\documentclass{aa}
\usepackage[pdftex]{graphicx}
\usepackage{epsfig}
\usepackage{amsmath}
\usepackage{wasysym}
\usepackage{txfonts}
\pdfoutput=1
\usepackage[pdftex]{color,xcolor}
\usepackage[pdftex]{graphicx}
\usepackage[backref]{hyperref}               
\usepackage{hyperref}
\hypersetup{pdfauthor=Paul Molliere}
\hypersetup{backref=true, pagebackref=true, hyperindex=true, breaklinks=true,colorlinks=true,urlcolor=blue, linkcolor=blue,citecolor=blue,pagecolor=red, bookmarks=true, bookmarksopen=true}
\usepackage{epstopdf}
\hypersetup{backref=true}
\usepackage{lipsum}
\usepackage{natbib}
\usepackage{mathtools}

\bibpunct{(}{)}{;}{a}{}{,} 
\usepackage{mhchem}

\def\f1{f_{\rm I}}
\def\mj{{\rm M}_{\textrm{\tiny \jupiter }}}

\newcommand{\rj}{{\rm R}_{\textrm{\tiny \jupiter}}}

\def\beq{\begin{equation}}
\def\eeq{\end{equation}}

\def\t2{\tau_{\rm II}}

\def\sigmas0{\Sigma_{\rm s,0}}

\def\({\left(}
\def\){\right)}
\def\<{\left<}
\def\>{\right>}

\newcommand{\rch}[1]{{#1}}

\begin{document}

\title{Detecting isotopologues in exoplanet atmospheres using ground-based high-dispersion spectroscopy}

\author{P. Molli\`{e}re\inst{1}  \and I.A.G. Snellen\inst{1}}

\institute{Leiden Observatory, Leiden University, Postbus 9513, 2300 RA Leiden, The Netherlands}

\offprints{Paul MOLLIERE, \email{molliere@strw.leidenuniv.nl}}

\date{Received 31 August 2018 / Accepted 19 December 2018}

\abstract
{The cross-correlation technique is a well-tested method for exoplanet characterization, having lead to the detection of various molecules, to constraints on atmospheric temperature profiles, wind speeds, and planetary spin rates. A new, potentially powerful application of this technique is the measurement of atmospheric isotope ratios. In particular D/H can give unique insights into the formation and evolution of planets, and their atmospheres.}
{In this paper we aim to study the detectability of molecular isotopologues in the high-dispersion spectra of exoplanet atmospheres, to identify the optimal wavelengths ranges to conduct such studies, and to predict the required observational efforts - both with current and future ground-based instrumentation.}
{High-dispersion (R=100,000)  thermal emission spectra, and in some cases reflection spectra, were simulated by self-consistent modeling of the atmospheric structures and abundances of exoplanets over a wide range of effective temperatures. These were synthetically observed with a telescope equivalent to the VLT and/or ELT, and analysed using the cross-correlation technique, resulting in signal-to-noise ratio predictions for the $^{13}$CO, HDO, and CH$_3$D isotopologues.}
{We find that for the best observable exoplanets, $^{13}$CO is well in range of current telescopes. \rch{We predict it will be most favorably detectable at 2.4 microns, just longward of the wavelength regions probed by several high-dispersion spectroscopic observations presented in the literature.} \rch{CH$_3$D can be best targeted at 4.7 microns, and may be detectable using 40m-class telescopes for planets below 600~K in equilibrium temperature. In this case, the sky background becomes the dominating noise source for self-luminous planets. HDO is best targeted at 3.7 microns, and is less affected by sky background noise. 40m-class telescopes may lead to its detection for planets with $T_{\rm equ}$ below 900~K. It could already be in the range of current 8m-class telescopes in the case of quenched methane abundances.} Finally, if Proxima Cen b is water-rich, the HDO isotopologue could be detected with the ELT in $\sim$1 night of observing time in its reflected-light spectrum.}
{Isotopologues will soon be a part of the exoplanet characterisation tools. Measuring D/H in exoplanets, and ratios of other isotopes, could become a prime science case for the first-light instrument \emph{METIS} on the European ELT, especially for nearby temperate rocky and ice giant planets. This can provide unique insights in their history of icy-body enrichment and atmospheric evaporation processes.}
\keywords{methods: numerical -- planets and satellites: atmospheres -- radiative transfer}
\titlerunning{Detecting isotopologues in exoplanet atmospheres}
\authorrunning{P. Molli\`ere \& I. Snellen}

\maketitle

\section{Introduction}
\label{sect:intro}

The cross-correlation technique is a well established tool for detecting the presence of molecular absorbers in exoplanet atmospheres. In addition, it can constrain a planet's atmospheric temperature profile, planetary spin rate and wind patterns, as well as its mass $-$ in the case of a non-transiting planet. Examples are the detection of CO in HD~209458~b \citep[which also constrained the planet's wind speed, see][]{snelleldekok2010}, CO in $\beta$ Pic~b  \citep[which also constrained the planet's spin rate, see][]{snellenbrandl2015}, CO in $\tau$~Bo\"otis~b \citep{brogisnellen2012,rodlerlopez2012}, H$_2$O in 51~Peg~b \citep{birkbydekok2017}, H$_2$O in  HD~88133~b and Ups~And~b \citep{piskorzbenneke2016,piskorzbenneke2017}, H$_2$O and CO in HD~179949~b \citep{brogidekok2014}, H$_2$O \citep{birkbykok2013} and CO in HD~189733b \citep{rodlerkuerster2013,kokbrogi2013}, and TiO in the temperature-inverted part of WASP-33b's atmosphere \citep{nugrohokawahara2017}. \citet{bryanbenneke2018} measured the rotation rates for a set of planets and brown dwarfs by cross-correlating with atmospheric models containing, among others, H$_2$O and CO opacities. Recently, atomic and ionized species were detected using cross-correlation techniques in the optical transmission spectrum of the hottest known planet Kelt-9b \citep{hoeijmakersehrenreich2018}. \rch{Also the techniques for extracting physical parameters from observations are improving, with the first study of high-resolution spectral retrieval presented recently by \citep{brogiline2018}.}

In the study presented here, we have investigated how the cross-correlation technique may be used to identify isotopologues in planetary atmospheres. Isotopologues are molecular chemical species with different numbers of neutrons in the nuclei of their constituent atoms. We have studied the feasibility of such a detection using current and next-generation instruments, in particular \emph{CRIRES+} \citep{follertdorn2014} on the Very Large Telescope, and \emph{METIS} \citep{brandlfeldt2014} on the ELT.  

Most of the near-infrared (NIR) molecular opacity of (hot) Jupiters originates from H-, C- and O- bearing species. In this study, we therefore concentrate  on the isotopologues of CO, H$_2$O and CH$_4$. H$_2$O and CH$_4$ are expected to be abundant in temperate, low-mass planets, with H$_2$O playing a key role in planet habitability. Isotopologue detections and subsequent measurements of isotop(ologu)e-ratios from these species can provide interesting insights in planet formation and atmospheric processes. In principle, this requires benchmarking the inferred ratios with those observed in the host star or the surrounding interstellar medium. However, although there exists scatter in our local neighborhood ($\sim$ 1 kpc), the carbon, oxygen and hydrogen isotope ratio variations are usually not larger than a factor of a few \citep[see, e.g.][]{milramsavage2005,polehamptonbaluteau2005,linskydraine2006}.\footnote{We note that for carbon and oxygen the isotopic ratios $^{12}$C/$^{13}$C, $^{16}$O/$^{17}$O and $^{16}$O/$^{18}$O decrease toward the galactic center \citep[see, e.g.,][]{milramsavage2005,romanomatteuci2017}. This is thought to be caused by dredge-up of heavier C and O isotopes as reaction-intermediates of the CNO cycle in AGB stars, which are later ejected into the ISM with the stars' outer layers.}
Hence, if exoplanet isotope ratios are found to be very different from the average local ISM values, this would make the planet stand out already without having to compare to the isotope ratios of its host star.

Typical isotope ratios in the neighborhood of the sun are $^{12}$C/$^{13}$C$\sim$\rch{70} \citep{milramsavage2005}, $^{16}$O/$^{17}$O$\sim$1600 \citep{romanomatteuci2017}, $^{16}$O/$^{18}$O$\sim$400 \citep{polehamptonbaluteau2005}, and deuterium to hydrogen (D/H) $\sim 2\times 10^{-5}$ \citep{linskydraine2006,asplund2009}. We note that in the solar system the oxygen and carbon isotope ratios are $\sim$30~\% larger, probably because the sun formed 4.5 Gyr ago, conserving the higher isotope ratios from that period \citep{claytonnittler2004,ayreslyons2013}.

Variations in the isotope ratios may provide clues on how planets form from the condensed and gaseous material in proto-planetary disks. In the solar system, such variations are most prominent in D/H \citep[see Figure 1 of][]{altweggbalsiger2015}. Volatile-rich primitive meteorites are found to be enriched in D by about a factor six  compared to the proto-solar nebula. This enrichment is found to be comparable or higher for Jupiter-family comets, and to be a factor 10$-$20 for Oort-cloud comets. \rch{We note, however that significant scatter exists, and the D/H value found in the Jupiter family comet 67P/GC is actually consistent with the highest D/H values found in Oort cloud comets, see \citet{altweggbalsiger2015}.} This trend of enrichment in D within the condensed volatiles appears to increase with distance from the Sun. This may be explained by a temperature-dependent fractionation of D into water ice \citep[see, e.g.,][]{geissreeves1981}, but we note that the Oort comets may have formed in the inner Solar System, and then been scattered outward by the giant planets \citep[see, e.g.,][and the references therein]{morbidelli2005}.

The gas giants Jupiter and Saturn are found to have D/H values of 
 $(2.6\pm 0.7)\times 10^{-5}$ and ${1.7}^{+0.75}_{-0.45}\times 10^{-5}$ respectively \citep{mahaffydonahue1998,lellouchbezard2001}, roughly in line with the proto-solar value ($2\times 10^{-5}$). In contrast, the ice giants Uranus and Neptune have measured ratios of $(4.4\pm 0.4)\times 10^{-5}$ and $(4.1\pm 0.4)\times 10^{-5}$ \citep{feuchtgruberlellouch2013}, hence they are D-enriched by a factor of approximately two. This is thought to be caused by atmospheric contamination by icy planetesimals. While the fraction of such contamination is low for Jupiter and Saturn, having no effect on their D/H values, this is significant for Uranus and Neptune.  Assuming that these icy planetesimals had an intrinsic D/H enrichment of an order of magnitude, similar to that of comets  (i.e. D/H $\sim2\times 10^{-4}$), and assuming an atmospheric enrichment as indicated by \citet{guillotgautier2014} for Uranus and Neptune, this can have caused the overall D-enrichment of their atmospheres by a factor of two. 
 
However, it is not yet known whether this scheme generally applies to planet formation processes. Numerous studies address atmospheric composition (relative to H--He), and its connection to a planet's formation history, often assuming that it is governed by gas that the planet accretes, instead of planetesimals \citep{obergmurray-clay2011,ali-dibmousis2014,thiabaudmarboeuf2014,hellingwoitke2014,marboeufthiabaud2014a,marboeufthiabaud2014b,madhusudhanamin2014,mordasinivanboekel2016,obergbergin2016,madhubitsch2016,cridlandpudritz2016}. Among these \citet{mordasinivanboekel2016} is a notable exception, assuming planetesimal, rather than gas enrichment.  In the case of gas enrichment, volatiles in the gas phase of the disk are expected to be partly sequestered into condensates, resulting in lower metallicities but unchanged isotopologue ratios in the disk gas. Hence, it is expected that isotopologue ratios in the atmospheres of extrasolar planets will not be significantly different from that of their host stars or local ISM, except if they are strongly contaminated by icy planetesimals, which have increased D/H values. The inferred trends of increasing planetary bulk metallicity as a function of decreasing planetary mass may be seen as a sign of planetesimal and solid body enrichment dominating over gas enrichment \citep{millerfortney2011,thorngrenfortney2016}. Recently, synthetic planet formation calculations by \citet{marboefthiabaud2018} have been able to reproduce the planetary mass -- D/H correlation in the Solar System, when applying the envelope pollution by planetesimals as advocated by \citet{mordasinivanboekel2016}.

In contrast to deuterium and hydrogen, there is no evidence for strong fractionation of either oxygen or carbon.  The $^{16}$O/$^{18}$O values inferred from comets such as 1P/Halley, 67P/Churyumov-Gerasimenko and C/2014 Q2 (Lovejoy) \citep[see][and the references therein]{altweggbockelee2003,altweggbalsiger2015,bivermoreno2016} are broadly consistent with the \rch{Solar System abundances}. The same holds for the oxygen isotopic abundances of primitive and differentiated meteorites \citep[systematic variations do exist, but only of the order of single digit percentage values, see, e.g.,][]{clayton1993,yurimotokuramoto2007}. Also, the $^{12}$C/$^{13}$C ratios observed in 11 different comets \citep[see][and the references therein]{altweggbockelee2003,bivermoreno2016} and chondrites of different types \citep[see, e.g.,][]{halboutmayeda1986,pearsonsephton2006} are all consistent with the \rch{Solar System abundances}. This means that neither C nor O condensates have a significant preference for a certain isotope, resulting in the gas and condensate phase C and O isotope ratios to be unaffected, at least down to a percentage level \citep[e.g.,][]{clayton1993,yurimotokuramoto2007}.
 
A process that can further affect the D/H value is atmospheric escape: thermal or ion pick-up escape processes preferentially retain more massive isotopologues \citep[see, e.g.,][]{jankoskyphillips2001}. Hence D is expected to be relatively more abundant than H in planets which have undergone significant (non-hydrodynamic) atmospheric loss. This seems to be indeed the case for Mars, which exhibits a D/H value of at least five to seven times larger than the terrestrial value \citep[see, e.g.,][]{krasnopolsky2015,villanuevamumma2015,clarkemayyasi2017}. The primordial value inferred for Mars is about twice the terrestrial value \citep{leshin2000}. Identifying water as the major hydrogen carrier, the above values correspond to equivalent ocean depths between at least 60 and 130 m having been lost from the Martian surface \citep{krasnopolsky2015,villanuevamumma2015}. For Venus the process of evaporation may have been even more extreme, because its atmospheric D/H is about 1000 times higher than the protosolar value of $2 \times 10^{-5}$ \citep{kulikovlammer2006}. 

There is also a particularly interesting prospect that isotopologue measurements could shed light on biological processes and help to provide evidence for the presence of extraterrestrial life. 
Living matter on Earth tends to favor  $^{12}$C over $^{13}$C when building organic carbon molecules: for organic carbon compounds the $^{13}$C/$^{12}$C ratio is 3~\% lower than for inorganic compounds \citep[see, e.g.,][and the references therein]{langmuirbroecker2012}. For Earth, the absolute $^{13}$C/$^{12}$C values in the organic and inorganic reservoirs depend on the fractionation of carbon between these two reservoirs, as well as on the total $^{13}$C/$^{12}$C (crustal) average. A possible tracer of life would thus be to compare the $^{13}$C/$^{12}$C ratios of atmospheric methane \citep[mostly organic origin on Earth, see, e.g.,][]{quayking1991} and CO$_2$ \citep[inorganic origin, if not derived from burning fossil fuels, see, e.g.,][]{ghoshbrand2003}. Alternatively, if the $^{13}$C/$^{12}$C crustal average of a terrestrial exoplanet was known, and the $^{13}$C/$^{12}$C in its oxidized (e.g. CO, CO$_2$) or reduced (CH$_4$) atmospheric components could be measured, then the carbon fractionation into organic (i.e. living, or recently living) matter could be inferred. However, assuming these processes would be the same as on Earth, these effects are very difficult to measure, requiring a precision in  absolute $^{13}$C abundance of $\sim$10$^{-4}$. For exoplanets, we expect this to be out of scope of any present or future planned instrument or telescope. 

In this paper, we focus on the detectability of carbon monoxide ($^{13}$CO), methane (CH$_3$D) and water (HDO), because we expect ($^{13}$CO) the least difficult to measure, and CH$_3$D and HDO to bear the greatest significance when seeking to probe a planet's formation and evolution history. In Section \ref{sect:model_descr} we describe how the planetary high-resolution spectra are modeled, and how the observations are simulated. In Section \ref{sect:detect_CO} we present our calculations for the detectability of $^{13}$CO in hot Jupiters, as a function of wavelength. In Section \ref{sect:hdo_seld_lum} we show how HDO may be found in self-luminous \rch{and irradiated} planets, as a function of effective temperature, with and without methane quenching in the atmosphere. In Section \ref{sect:ch3d_self_lum} we study the detectability of CH$_3$D, with the same planetary setup. Finally, the case of HDO in terrestrial exoplanet atmospheres is studied in Section \ref{sect:proxima_hdo}, where we assume a twin Earth as an input model for Proxima~Cen~b. Our results are summarised and discussed in Section \ref{sect:discussion_summary}.

\section{Model description}
\label{sect:model_descr}

\subsection{Atmospheric structure}
\label{sect:atmo_struct}

The atmospheric temperature and abundance structures used to generate the high resolution spectra are derived from self-consistent atmospheric models. In the calculations presented here, structures are obtained with \emph{petitCODE} \citep{mollierevanboekel2015,mollierevanboekel2017}, except for the results shown in Section \ref{sect:proxima_hdo}.  \emph{petitCODE} calculates the atmospheric structures of exoplanets in 1-d  in radiative-convective and chemical equilibrium. The radiative transfer considers both absorption and scattering processes. Only gas opacities are considered in the calculations presented here, but clouds can optionally be included in \emph{petitCODE} calculations, in a self-consistent fashion. The gas opacity species considered here are H$_2$O, CO, CO$_2$, OH \citep[HITEMP, see][]{rothman2010}, CH$_4$, NH$_3$, PH$_3$, HCN \citep[ExoMol, see][]{tennyson:2012aa}, as well as H$_2$, H$_2$S, C$_2$H$_2$ \citep[HITRAN, see][]{rothman2013}, Na , K \citep[VALD3, see][]{piskunov1995}, and CIA of H$_2$--H$_2$ and H$_2$--He \citep{borysowfrommhold1989a,borysowfrommhold1989b,richardgordon2012}. For H$_2$ and CO, also the UV electronic transitions by \citet{kurucz1993} are included, as well as the Rayleigh scattering opacities for H$_2$, He, CO$_2$, CO, CH$_4$ and H$_2$O. For the cross-sections, 
the values reported in \citet{dalgarnowilliams1962} (H$_2$),
\citet{chandalgarno1965} (He), \citet{sneepubachs2005} (CO$_2$, CO, CH$_4$) and \citet{harveygallagher1998} (H$_2$O) are used.

\emph{petitCODE} is a well-tested tool for calculating exoplanet atmospheric structures, and has recently been benchmarked against the \emph{ATMO} \citep{tremblinamundsen2015} and \emph{Exo-REM} \citep{baudinobezard2015} codes \citep{baudino:2017aa}. It was used for a parameter study of irradiated atmospheres \citep{mollierevanboekel2015}, and for generating predictions of exoplanet observations with the James Webb Space Telescope (\emph{JWST}) for high-priority targets \citep{mollierevanboekel2017}. Moreover, \emph{petitCODE} enabled the atmospheric characterization of the self-luminous planet {51 Eri b} \citep{samlandmolliere2017}, and constrained the atmospheric properties of several transiting exoplanets \citep{mancinikemmer2016, mancinigiordano2016, mancinisoutworth2017, southworthmancini2016}. Finally, it connected planet formation models with synthetic atmospheric observations in \citet{mordasinivanboekel2016}. 

\subsection{High resolution spectra}
\label{sect:high_res_code}

The high resolution spectra were calculated with a new radiative transfer code, based on \emph{petitCODE}, that we report on here for the first time. It uses the same molecular opacity database as \emph{petitCODE}, in which pressure and temperature-dependent opacities are stored at a resolution of $\nu/\Delta\nu = 10^{6}$. The atmospheric structure calculations in \emph{petitCODE} are carried out at a lower resolution, making use of the correlated-k approximation \citep{goody1989,lacis_oinas1991,fuliou1998}, as described in Appendix B of \citet{mollierevanboekel2015}. The new high resolution radiative transfer code presented here uses the opacity database of \emph{petitCODE} at its intrinsic resolution, and in a line-by-line, rather than a correlated-k treatment. Identical to the capabilities of \emph{petitCODE}, both transmission and emission spectra can be calculated. Because the work presented in this paper focuses on the high-resolution NIR to MIR emission spectra of cloudless atmospheres, scattering is currently neglected, but scattering is included in the atmospheric structure calculations, as described in Section \ref{sect:atmo_struct} above.

\subsection{Synthetic observations}
\label{sect:synth_obs}

For a given set of planet--star parameters (stellar effective temperature and radius, planetary semi-major axis, radius, mass and atmospheric composition) self-consistent atmospheric structures are calculated, and used for generating the high-resolution emission flux $F_{\rm Planet}(\nu)$ in the planet's rest frame, where $\nu$ denotes the frequency.
In order to generate the planet's signal as it would be seen by an instrument on Earth, the rest frame frequency values are first shifted according to $\nu \mapsto \nu(1- v_{\rm rad}/c)$, where $v_{\rm rad}$ is the radial velocity between the planet and the observer, and $c$ is the speed of light. Subsequently, the host star's flux is included by adding a flat white spectrum to the planetary flux,
\beq
F_{\rm tot}(\nu) = F_{\rm Planet}(\nu) + F_*,
\eeq
where the contrast between planet and star is a free parameter. We assume that the stellar lines can be perfectly removed during the data analysis. A second free parameter is the signal-to-noise-ratio (SNR) of the stellar flux, SNR$_*$. For convenient numerical modeling, we set $F_*={\rm SNR}_*^2$ (assuming Poisson noise), and replace $F_{\rm Planet}(\nu)$ with $c\cdot {\rm SNR}_*^2\cdot F_{\rm Planet}(\nu)/\bar{F}_{\rm Planet}$, where $\bar{F}_{\rm Planet}$ is the average planetary flux and $c$ is the desired average contrast in the considered wavelength region. Subsequently, the total flux  $F_{\rm tot}(\nu)$ is multiplied with a transmission model for the Earth's atmosphere, $\mathcal{T}(\nu)$, such that the total flux that reaches the ground-based telescope is
\beq
F_{\rm tel}(\nu) = \mathcal{T}(\nu)F_{\rm tot}(\nu).
\eeq
Subsequently, the spectrum is convolved to the intrinsic spectral resolution of the instrument, and binned to the wavelength steps of the instrument. Both the instrument resolution and the number of pixels per resolution element are free parameters. We follow the convention of defining the instrumental resolution $\nu/\Delta \nu$ such that $\Delta \nu$ is the full width half maximum (FWHM) of the line spread function (LSF) of the instrument's dispersing element. We assume a Gaussian distribution for the LSF, hence the relation between its standard deviation and the instrument resolution is $\Delta \nu = 2\sqrt{2 \ {\rm ln}2}\ \sigma$. In the following, the result of convolving and re-binning $F_{\rm tel}(\nu)$ will be denoted $\tilde{F}_{\rm tel}$, where $\nu'$ is the frequency corresponding to the instrument pixel,
with the uncertainty being
\beq
\sigma_{\rm tel}(\nu') = \sqrt{\tilde{F}_{\rm tel}(\nu')}.
\eeq
The final simulated observation $F_{\rm obs}(\nu')$ is obtained by perturbing $\tilde{F}_{\rm tel}$ with a Gaussian with standard deviation equal to $\sigma_{\rm tel}$.

\begin{table}[t]
\centering
\begin{tabular}{cccc}
Parameter & Value & Reference \\ \hline \hline
$T_*$  & 6260 K & SI04 \\
$R_*$ & 1.14  $\rm R_\odot$  & SI04 (a) \\
$\rm [Fe/H]_{*}$ & 0.22  & SI04 \\
$d$ & 0.045  AU & WM07 \\
$M_{\rm Planet}$ & 0.98  $\mj$  & BdK14 \\
$R_{\rm Planet}$ & 1.4  $\rj$  & (b) \\
\rch{$T_{\rm equ}$} & \rch{1519~K} & \rch{$T_*$, $R_*$, $d$}\\
Orbital inclination $i$ & 67.7$^\circ$ & BdK14 \\
Line absorbers & H$_2$O, CO$_2$, CO, CH$_4$ & (c) \\
& HCN, H$_2$S, NH$_3$, H$_2$, & \\
& PH$_3$, C$_2$H$_2$, OH, Na, K & \\
Rayleigh scatterers & H$_2$, He & (c) \\
CIA & $\rm H_2$--$\rm H_2$, $\rm H_2$--$\rm He$ & (c) \\
Elemental abundances & $\rm [Fe/H]_{\rm Planet} = 0.83$ &  (d) \\
Chemical abundances & chemical equilibrium & (e) \\
\hline \hline
\end{tabular}
\caption{Parameters used for the self-consistent structure calculations for HD~179949b. (a): the stellar radius was inferred using the stellar mass and surface gravity reported in \citet{santosisraelian2004}. (b): HD~179949b is a non-transiting planet, hence its radius is unknown. The value chosen here corresponds to the radii typically found for  planets of that mass and insolation strength (see, e.g., \url{http://exoplanets.org}) (c): the references for the line opacity database of \emph{petitCODE} can be found in Section \ref{sect:atmo_struct}. (d): for the solar abundances we assumed the values reported by \citet{asplund2009}. For the planet's atmospheric enrichment the solar abundances were scaled with a value that was chosen following the method described in Section 4.1 of \citet{mollierevanboekel2017}. (e): we used the chemical equlibrium code described in \citet{mollierevanboekel2017}. References: BdK14: \citet{brogidekok2014}; SI04: \citet{santosisraelian2004}; WM07: \citet{wittenmyerendl2007}}
\label{tab:tau_boo_params}
\end{table}

\section{$^{13}$CO in hot Jupiter atmospheres}
\label{sect:detect_CO}

The prescription for generating simulated observations described above is used for a case study of $^{13}$CO in a hot Jupiter atmosphere. The HD~179949b system was chosen as a benchmark. It hosts a non-transiting gas giant \citep[discovered by][]{tinneybutler2001} with an equilibrium temperature of $T_{\rm equ}=1519$~K (see Table \ref{tab:tau_boo_params}) . Ground-based high-dispersion spectroscopic observations with \emph{CRIRES} on ESO's Very Large Telescope (VLT) have already shown the presence of both water (${\rm SNR}=3.9$) and CO (${\rm SNR}=5.8$) in the planet's atmosphere \citet{brogidekok2014}. 

\subsection{Synthetic HD~179949b observations}
\label{sect:hd179modelobs}

A self-consistent atmospheric structure was calculated as described in Section \ref{sect:atmo_struct}, assuming the input parameters from Table~\ref{tab:tau_boo_params}. Synthetic observations consisting of 
100 high-resolution observations in orbital phase from $-45^\circ$ to $+45^\circ$ around superior conjunction were generated as described in Sections \ref{sect:high_res_code} and \ref{sect:synth_obs}, taking into account the appropriate Doppler shifts of the planetary spectrum due to the radial component of its orbital velocity, and the waxing and waning of the planet. Also, the geometric effect of the orbital inclination on the area of the visible dayside was included. The flux of the visible part of the dayside was assumed to be uniform and equal to the average dayside flux obtained from \emph{petitCODE}, and to be zero on the nightside. For the high resolution spectra, only the line opacities of H$_2$O, CO$_2$ and CO were included, as well as H$_2$--H$_2$ and H$_2$--He CIA. For the planet-to-star contrast we used the values stemming from the self-consistent calculations.

The ${\rm SNR}_*$ as function of wavelength was calculated relative to the value at 2.3~$\mu$m following
\beq
{\rm SNR}_*(\lambda) = {\rm SNR}_*(2.3 \ {\rm \mu m})\cdot\sqrt{\frac{\left<F_*(\lambda)\right>}{\left<F_*(2.3 \ {\rm \mu m})\right>}}.
\label{equ:stellar_SNR_scaling}
\eeq
Here $\left<F\right>$ denote the average fluxes in the wavelength ranges of interest.
The telluric transmission model was generated using the \emph{ESO SkyCalc}\footnote{\url{https://www.eso.org/observing/etc/skycalc/}} tool \citep{nollkausch2012, jonesnoll2013}, assuming a stellar altitude of $60^\circ$ (airmass=2). The elevation was set to 2640 m, corresponding to the summit of Cerro Paranal, \rch{adopting also the median precipitable water vapor amount of Paranal (2.5 mm)}. An instrument resolution of $\nu/\Delta\nu = 10^5$ was assumed with wavelength steps corresponding to 3 pixels per resolution element $\Delta \nu$.

The CO isotopologue ratios were assumed to be the same as 
in the HITRAN/HITEMP databases (their \texttt{molparam.txt} file): $^{12}$C$^{16}$O constitutes 98.7 ~\%, while $^{13}$C$^{16}$O constitutes 1.1~\% of all CO molecules. The HITRAN/HITEMP values are based on the compilation of telluric isotopic abundances by \citet{debievregallet1984}. We note that in the case of  $^{12}$C/$^{13}$C, variations in the Solar System and its neighborhood are small (as discussed in Section \ref{sect:intro}), justifying the use of the telluric values, given in Table \ref{tab:nom_iso_abund}. 

\vspace{5mm}
\subsection{Analysis of the synthetic observations}
\label{sect:data_red}

\begin{figure}[t!]
\centering
\includegraphics[width=0.495\textwidth]{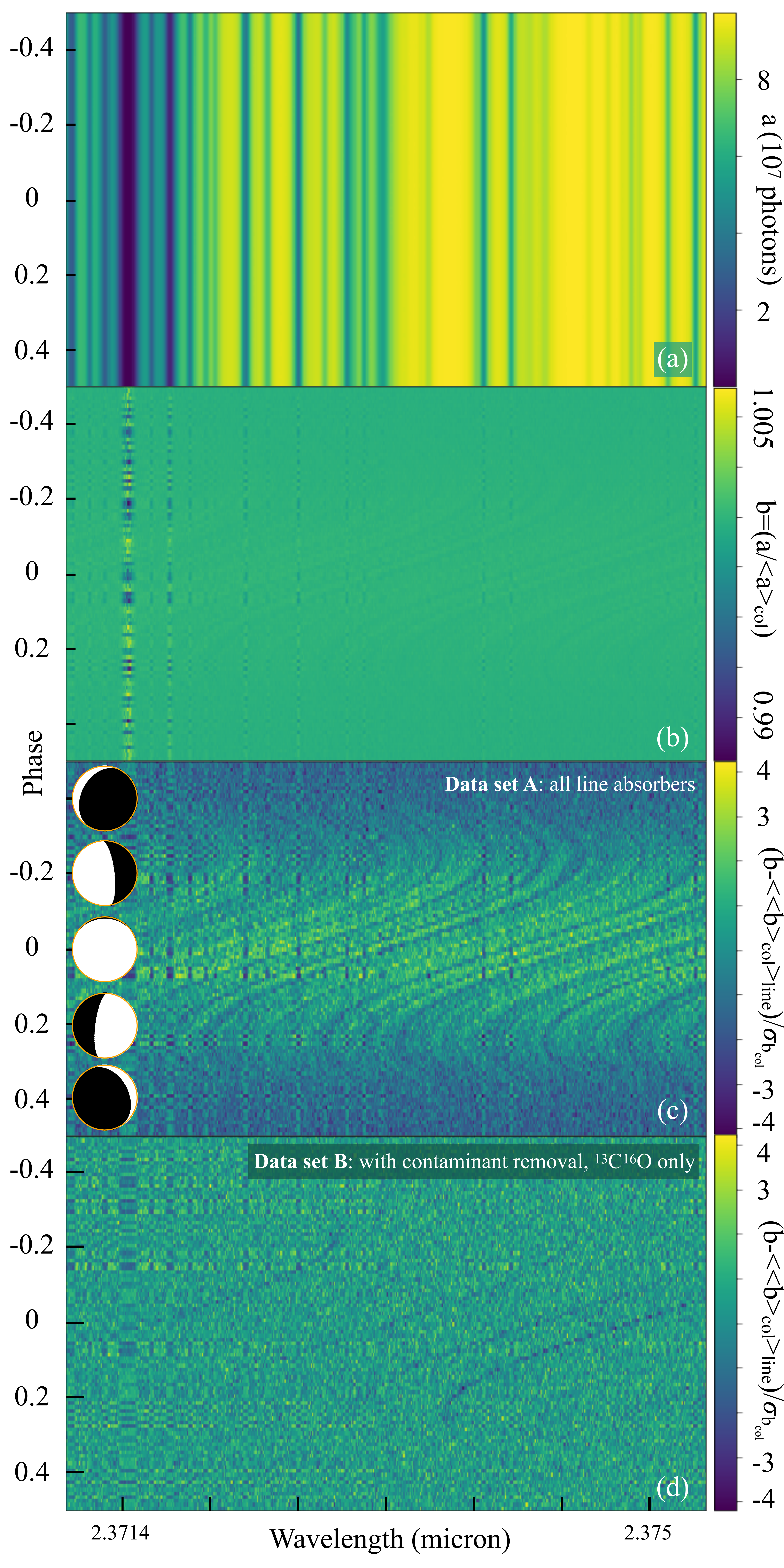}
\caption{Different analysis steps of the simulated observations. {\it Panel (a)}: raw synthetic \rch{observation (number of photons as a function of wavelength and phase)}; {\it Panel (b)}:  Data after the telluric correction\rch{, as described in the text}; {\it Panel (c)}: \rch{data of Panel (b) after subtracting the total wavelength and phase mean, and normalising each column by its standard deviation, to suppress noisy wavelength regions.} This is data set A; {\it Panel (d)}: The same as {\it Panel (c)}, but with the signal of all contaminant line absorbers removed, making the lines originating from $^{13}$C$^{16}$O visible. This is data set B.}
\label{fig:data_reduction}
\end{figure}

For the analysis of the synthetic observations, standard methods, as have been used for real high-dispersion observations, were applied \citep[see, e.g.,][]{brogisnellen2012,brogidekok2014}. The main steps of the analysis are briefly described below. The data \rch{(emission spectra of star and planet)} are organized as a two-dimensional matrix \rch{(see, e.g., uppermost panel of Figure \ref{fig:data_reduction})},  where the columns represent the wavelength steps, and rows the spectra taken at different orbital phases. 

\rch{The goal of the data reduction is to construct a nominal data set, containing the planet flux, from observations obtained with the procedure described in Section \ref{sect:synth_obs}. This will be called data set A, it is the planetary flux that an observer would measure, after a successful data reduction.} \rch{We also will create a data set B, which is identical to A, except that the lines of all spectroscopically-active molecules, except for $^{13}$C$^{16}$O, have been removed}: this is done by calculating planet spectra which contain all species, except the targeted $^{13}$C$^{16}$O isotopologue. Multiplying this new spectrum with the telluric transmission model\footnote{\rch{The telluric transmission model was measured from the synthetic data set A, to emulate a real data analysis as closely as possible, see below.}}, the new spectrum is then subtracted from \rch{the synthetic observation,  prior to reduction. After reduction, this will result in data set B. Hence, data set B only contains the lines of the targeted $^{13}$C$^{16}$O, at the correct relative line strengths as they are present in data set A.} Using data set B significantly simplifies the analysis of the contribution of the $^{13}$C$^{16}$O isotopologue in the spectra\rch{, because with it we can directly measure the signal strength imprinted onto the observation by $^{13}$C$^{16}$O alone.} \rch{This is different from the analysis of real observations, where }all available information on the planet spectrum, including those from other observations, will be used to constrain the atmospheric structure and relative volume mixing ratios of the relevant spectroscopically-active molecules \citep{brogiline2018}. Subsequently, spectra will need to be modeled assuming a range of isotope ratios and compared to the data. If the data are significantly better fitted by models for which the secondary isotopologue is included, this isotopologue is detected and an isotope-ratio can be inferred.
\begin{figure*}[t!]
\centering

\includegraphics[width=0.99\textwidth]{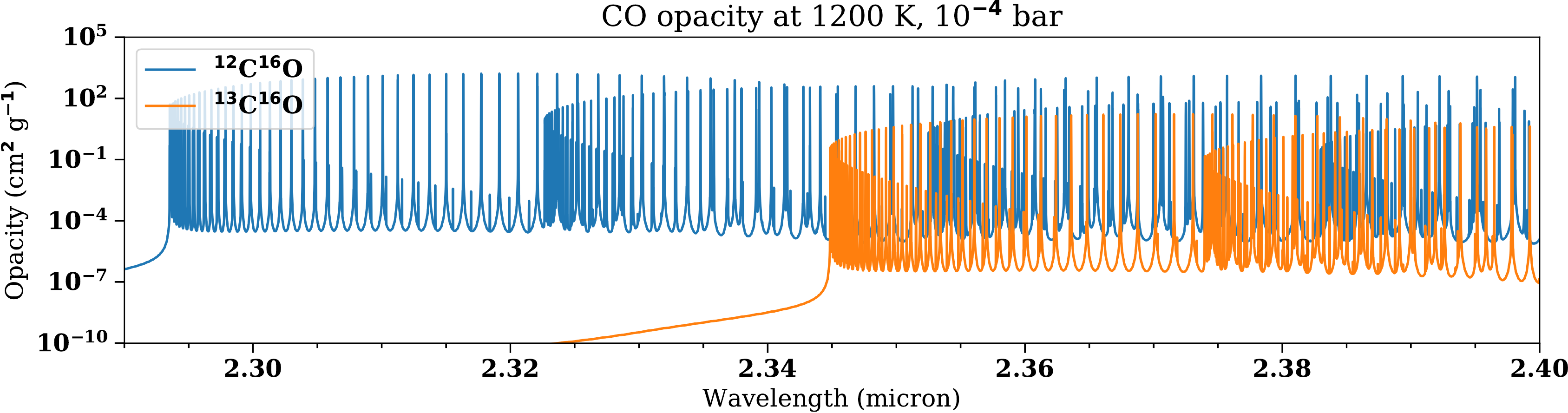}
\caption{Opacities of $^{12}$C$^{16}$O (blue) and $^{13}$C$^{16}$O (orange), shown at $T=1200$~K and \rch{$P=10^{-4}$~bar, which are representative values for the pressures and temperatures probed by CO lines at high resolution, in HD~179949b}. The opacities have been scaled such that $^{12}$C$^{16}$O constitutes 98.7~\% and $^{13}$C$^{16}$O 1.1~\% of all CO molecules.}
\label{fig:co_23_micron}

\vspace{3mm}

\includegraphics[width=0.495\textwidth]{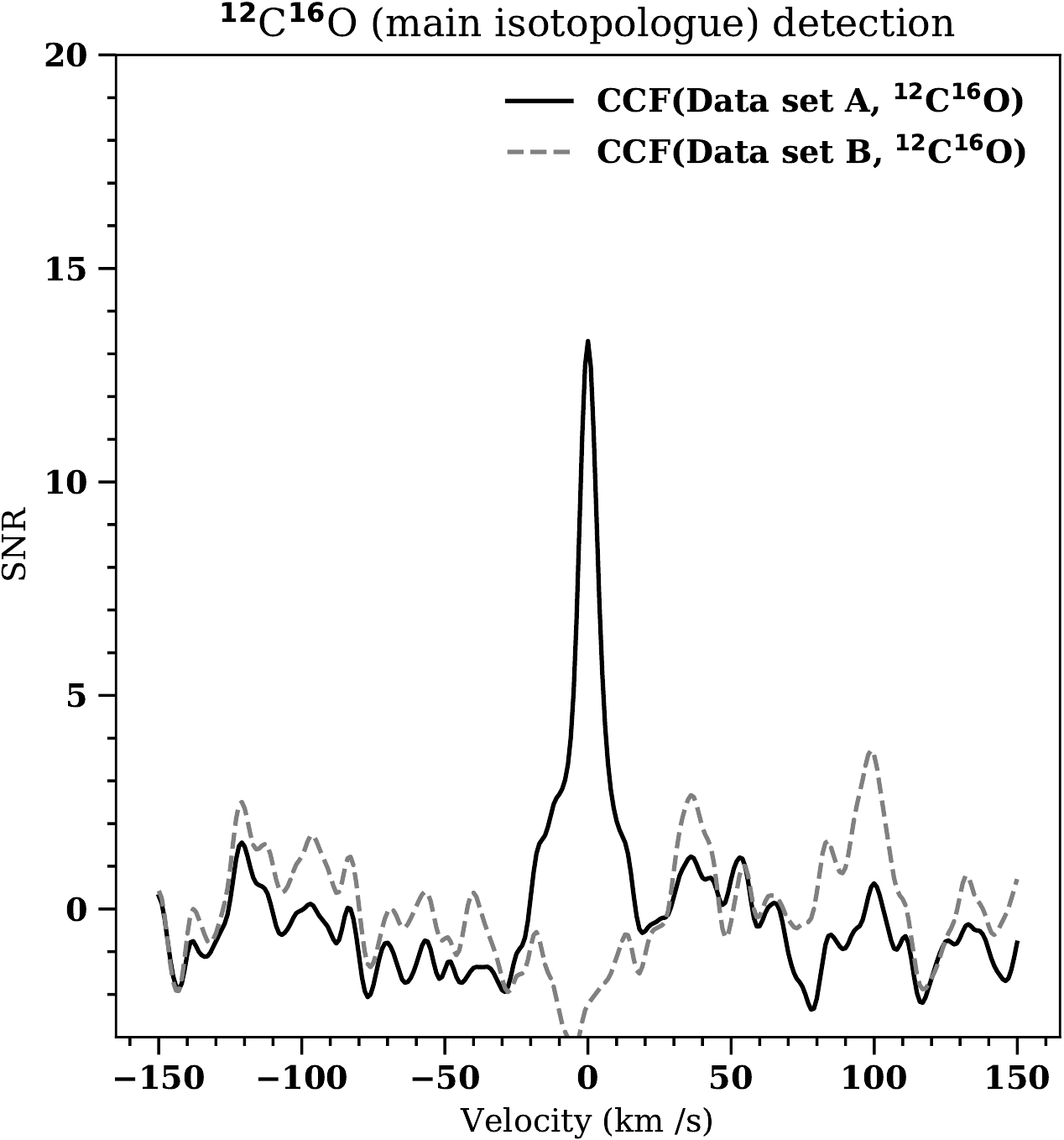}
\includegraphics[width=0.495\textwidth]{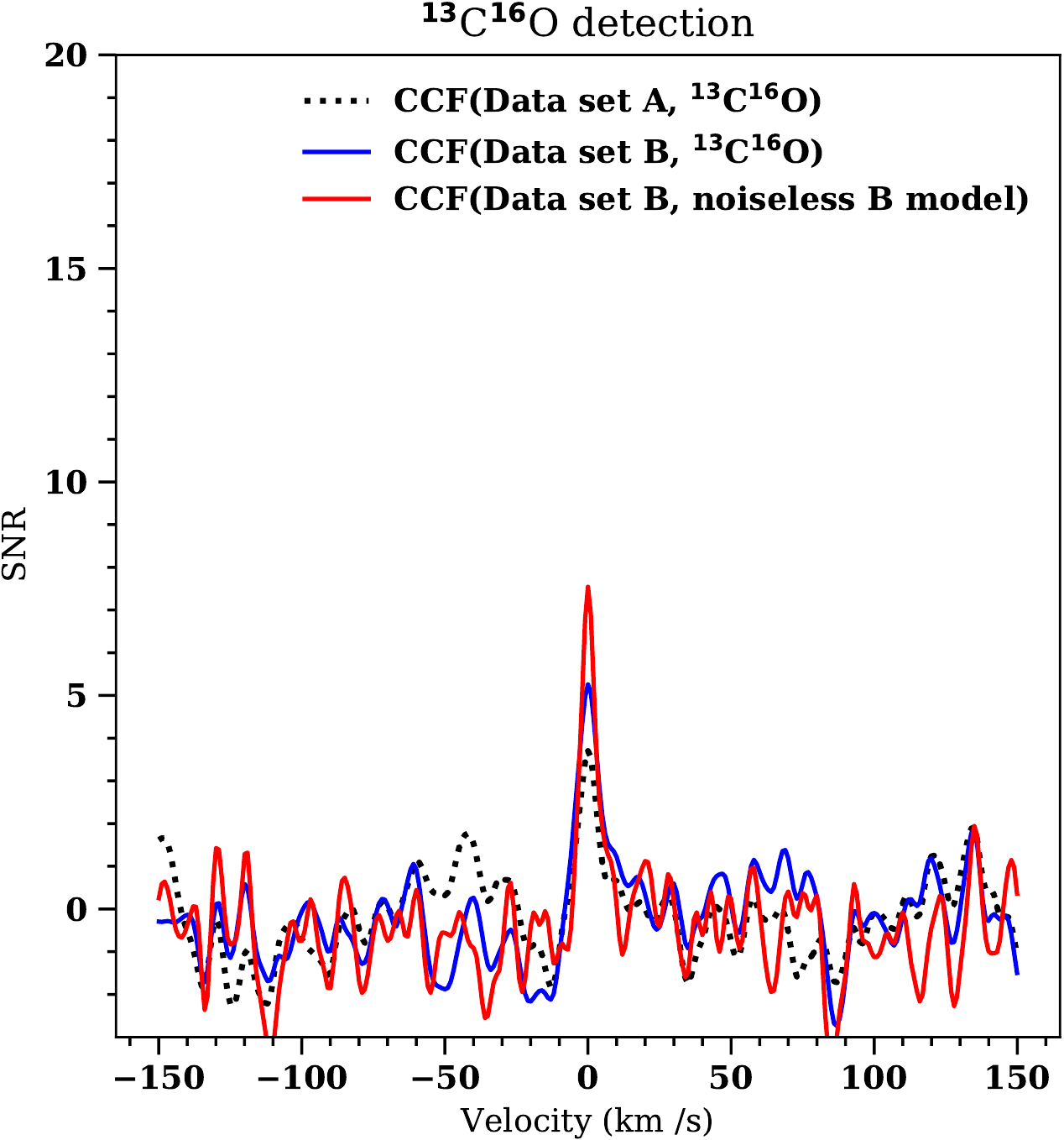}
\caption{Results from simulated observations of HD 179949\,b consisting of 100 spectra with an SNR$_*$ of 200 each, with a R=100,000 spectrograph covering the wavelength range $\lambda = 2.32 - 2.45   \ \mu$m.
{\it Left panel:} Cross-correlation function
(CCF) using a pure $^{12}$C$^{16}$O template spectrum on data set A (black solid line) which contains all species, and data set B (grey dashed line) which contains lines of $^{13}$C$^{16}$O only. Data set A gives a signal with an SNR of $\sim$13, while the main isotope, as expected, is not detected in data set B. 
{\it Right panel:} CCF using a pure $^{13}$C$^{16}$O template spectrum on data set A (black dotted line) and data set B (solid blue line). In addition, we show the CCF using a noiseless telluric-free model for data set B as a template (red solid line). The $^{13}$C$^{16}$O isotope is detected at an SNR of $\sim$3.5 and $\sim$5 in data set A and B respectively. By using the perfect noiseless input model as a template, $^{13}$C$^{16}$O's detection SNR is further increased to an SNR of $\sim$7. To reach the latter significance, the planet atmosphere needs to be well constrained.}
\label{fig:test_CCFs}
\end{figure*}
Hence, \rch{here }we do not study the retrievability of isotopologue abundances for atmospheres\rch{, instead we study the strength of the signal that a given isotopologue imprints on the observation. We will also show the $^{13}$C$^{16}$O detection SNRs, obtained from a classical analysis, when, for example, cross-correlating a pure $^{13}$C$^{16}$O model with data set A.} \rch{All the necessary procedures for the data reduction are described in the following two steps, which are identical for the data sets A and B.}

\rch{The synthetic observations are shown in Panel (a) of Figure \ref{fig:data_reduction}.} For every column the median value is calculated which will be used to normalise the data. This value as function of wavelength is the best estimate of the telluric absorption line spectrum, which in our simulations is kept constant throughout the observations, implying, within noise-limits, a perfect telluric subtraction. 
Panels (a) and (b) in Figure \ref{fig:data_reduction} show a small cutout of the simulated observations before and after the telluric correction. For clarity, an extremely high signal-to-noise of  ${\rm SNR}_*=10,000$ is used, with $c=4\times10^{-4}$, making the effects of the different analysis steps visible. 
The full orbital phase is shown to demonstrate the effect of the waxing and waning of the planet, but for the analysis below only phase angles varying between  $-45^\circ$ and $+45^\circ$ around superior conjunction are considered.

\rch{Finally, to obtain data set A, } the data shown in Panel (b) of Figure \ref{fig:data_reduction} is scaled by its standard deviation in each column, suppressing the parts of the data affected by strong telluric absorption. \rch{This is achieved by first calculating the total flux-average of all points in Panel (b) (over all columns and lines), and subtracting that value from all points in Panel (b). All points in a given column are then divided by their column-wide standard deviation.} Panel (c) in Figure \ref{fig:data_reduction} shows the simulated data after this step. \rch{This is data set A. This step is done} to prevent the cross-correlation signal to be dominated by the more noisy data. Panel (d) of Figure \ref{fig:data_reduction} shows the same as Panel (c), but now for data set B, hence showing only the lines of $^{13}$C$^{16}$O isotopologue.

\begin{figure*}[t!]
\centering
\includegraphics[width=1.0\textwidth]{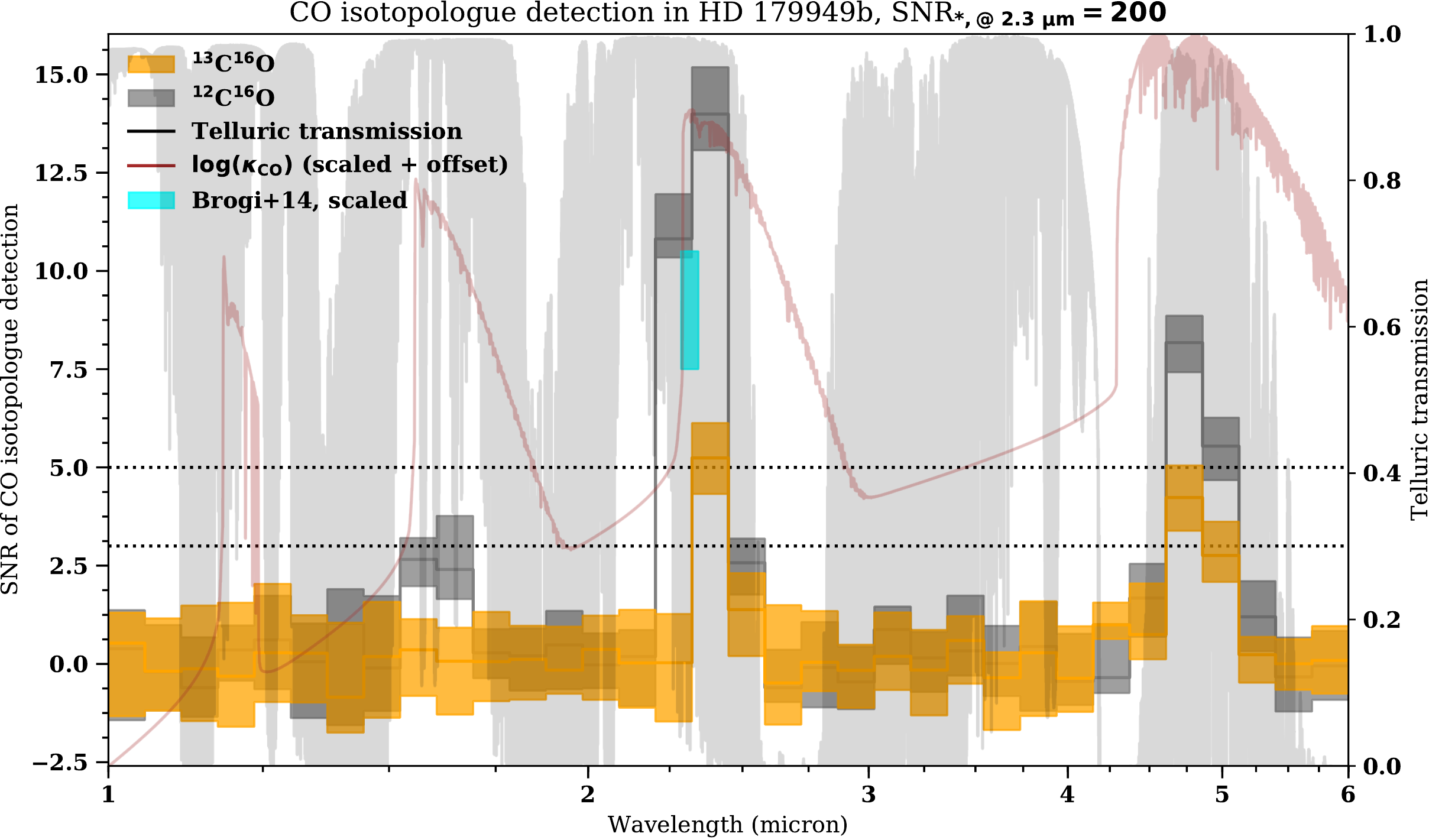}
\caption{Wavelength-dependent detection SNR for the secondary $^{13}$C$^{16}$O (orange boxes) and the main $^{12}$C$^{16}$O isotopologue (gray boxes). The box widths correspond to the wavelength range of the synthetic observations, while the height corresponds to the 16 and 84 percentiles of the measured SNRs, as derived by running the simulations multiple times. For $^{13}$C$^{16}$O the SNR of data set B cross-correlated with the pure isotopologue template spectrum are shown, which is conservative, corresponding to the solid blue line in Figure \ref{fig:test_CCFs}. 
For $^{12}$C$^{16}$O the SNR of data set A cross-correlated with the pure isotopologue template spectrum are shown.
We assumed 100 observations with ${\rm SNR}_*(2.3 \ \mu{\rm m})=200$, used to calculate the stellar SNR as function of wavelength. In the background, we show the telluric transmission model (gray solid line), as well as the scaled and offset logarithm of the CO opacity at $T=1200 \ {\rm K}$ and $P=10^{-4} \ {\rm bar}$ (light red solid line).  The CO detection by \citet{brogidekok2014} is shown in cyan. The actual SNR value of the CO detection in \citet{brogidekok2014} is 5.8, but due to the larger wavelength coverage of our bins one has to scale this value up to a SNR of 9. This is somewhat lower than our prediction for $^{12}$C$^{16}$O, but the SNR of their observations is also smaller. }
\label{fig:wlen_study}
\end{figure*}

\subsection{Cross-correlation signal at 2.4 microns}
\label{ref:cross_corr_CO}

We first demonstrate the use of the cross-correlation technique to detect the $^{13}$C$^{16}$O isotopologue of carbon monoxide, considering a wavelength range of 2.32 to 2.45 microns, just redward of the wavelength regions probed by several previous observations targeting CO in hot-Jupiter atmospheres \citep[e.g.][]{brogidekok2014}. Since these observations only probed out to 2.345 microns, they just missed the band head of  $^{13}$C$^{16}$O. In Figure \ref{fig:co_23_micron}, we show the opacities of $^{12}$C$^{16}$O (blue) and $^{13}$C$^{16}$O (orange) between 2.29 and 2.40 microns. The $^{13}$C$^{16}$O band head at 2.345 microns is clearly visible. 

The 100 simulated high-dispersion spectra were each given a signal-to-noise of SNR$_*=200$ per wavelength step. First we consider the cross-correlation of data set A (which includes the lines from all molecules and isotopologues) with a pure $^{12}$C$^{16}$O template spectrum, \rch{assuming that the radial velocity curve of the planet is perfectly known}. For the \rch{pure $^{12}$C$^{16}$O template spectrum we assumed the same atmospheric structure as used for creating the full planetary spectrum, but} include only the $^{12}$C$^{16}$O opacity in the spectral calculation. This results in a cross-correlation signal with an SNR of $\sim$13 (see left panel of Figure \ref{fig:test_CCFs}). This is as expected. The planet-to-star contrast of this system is $7 \times 10^{-4}$ at 2.3 microns, and the SNR of the combined 100 spectra is $2\times10^{3}$, implying an SNR on the planet spectrum of $\sim1.4$ per wavelength step. This is very similar to precisions already reached with existing observations, albeit for a smaller and slightly blueward wavelength region \citep{brogidekok2014}. With on the order of 100 strong CO lines in the targeted wavelength region, this combines to an overall SNR of $\sim\sqrt{100}\cdot 1.4 = 14$ (see Appendix \ref{app:cross_corr_SNR_lines_equ_strength} for a derivation), which is in good agreement with the SNR resulting from our more detailed simulations. As a control, we also cross-correlated data set $B$ (from which all spectral lines of the main isotopologue were removed) in the same way, and naturally no signal was detected (see left panel of Figure \ref{fig:test_CCFs}).

Subsequently, we cross-correlated data set $B$, that is the observation from which all lines of molecules and isotopes other than $^{13}$C$^{16}$O have been subtracted, with two different models. First, with a pure template spectrum, containing lines only from $^{13}$C$^{16}$O. This is shown in the right panel of Figure \ref{fig:test_CCFs}. $^{13}$C$^{16}$O is the most abundant of the secondary isotopologues (1.1~\% of all CO molecules, see Table \ref{tab:nom_iso_abund}), and therefore studied here. This results in a cross-correlation signal with an SNR of $\sim$5. A better matching cross-correlation template is given by the difference between a full planet spectrum model minus the same constructed without the opacity of $^{13}$C$^{16}$O, a noiseless model B spectrum. This takes potential shielding of $^{13}$C$^{16}$O lines by the other isotopologues or molecules into account.  
This results in a cross-correlation signal with an SNR of $\sim$7, and is expected in the case that sufficient spectral information is available such that the planet atmosphere can be well modelled. Even in the worst case, that is without removal of the main isotopologue and other molecules from the data (using data set $A$), the signal from the $^{13}$C$^{16}$O, correlating with the pure $^{13}$C$^{16}$O template spectrum, is still detected at an SNR of $\sim$3.5, see right panel of Figure \ref{fig:test_CCFs}.

\subsection{Wavelength and SNR study}
\label{ref:wlen_SNR_co}

In the previous section we showed that $^{13}$C$^{16}$O should be readily detectable at 2.4 microns. Here we investigate how such detectability varies as function of wavelength. This is shown in Figure \ref{fig:wlen_study}, where the expected SNR of $^{13}$C$^{16}$O (and $^{12}$C$^{16}$O) for the same benchmark hot Jupiter is shown as function of wavelength, for blocks of $\lambda / \Delta \lambda$ of 20, at a resolving power of R=100,000, with a wavelength sampling of three pixels, a stellar ${\rm SNR}_*(2.3 \  {\rm \mu{\rm m}})$ of 200 per step per exposure, and 100 exposures. The stellar SNR at wavelengths different from 2.3 microns was obtained using Equation \ref{equ:stellar_SNR_scaling} (using \emph{PHOENIX} models \citep{hauschildtbaron1999b} for the host star, as described in \citealt{mollierevanboekel2015}). It changes from roughly 300 at 1 $\mu$m to 80 at 6 $\mu$m.

\rch{The sky background will rise steeply for ground-based observations at wavelengths longer than 4.2 microns. To study the importance of this effect we additionally added the wavelength-dependent photon noise of the sky to the noise budget. The sky radiance was obtained from the \emph{SkyCalc} tool, with the same settings as used for the telluric transmission model. Moreover, we assumed an 8m-class telescope with 0.1 total throughput, which resulted in an exposure time of 28 seconds for each of the individual 100 spectra, if requiring to reach ${\rm SNR}_*(2.3 \  {\rm \mu{\rm m}})=200$. The thermal emission of the telescope was not included.} The grey and orange lines indicate the SNR as function of wavelength for the detection of the $^{12}$C$^{16}$O and $^{13}$C$^{16}$O isotopologues, respectively.

For $^{13}$C$^{16}$O we show the SNR arising from correlating data set B with a pure $^{13}$C$^{16}$O spectral template, which is conservative, corresponding to the solid blue line in Figure \ref{fig:test_CCFs}.
\rch{When using the noise-free data set B spectrum as a cross-correlation template, the SNR at, for example, 4.7 $\mu$m would increase from $\sim$4.5 to $\sim$6.5.}
For the $^{12}$C$^{16}$O detection we show the SNR arising from correlating data set A with a pure $^{12}$C$^{16}$O spectral template. The simulations at each wavelength were run multiple times to reduce the stochastic scatter. In Figure \ref{fig:wlen_study} the 16 to the 84 percentiles of the SNR distributions are indicated as orange or gray boxes, corresponding to the 1$\sigma$ uncertainty ranges. The pure isotopologue models were chosen to cross-correlate with the data, to allow direct comparison to the CO detection by \citet{brogidekok2014}, who used pure (multi-isotopologue) CO models.

\begin{figure*}[t!]
\centering

\includegraphics[width=0.99\textwidth]{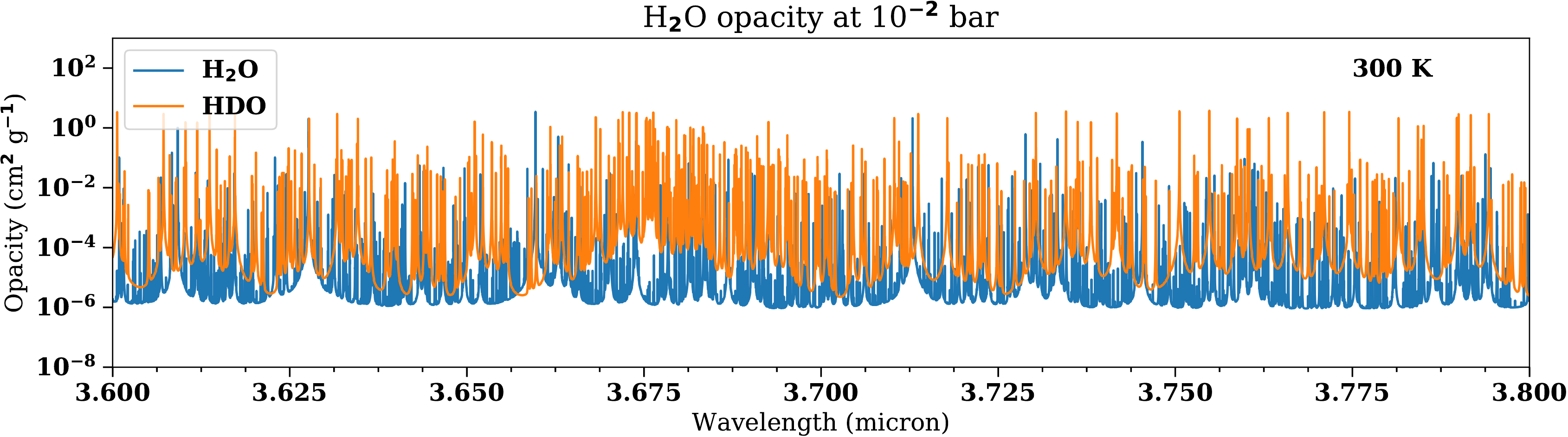}
\includegraphics[width=0.99\textwidth]{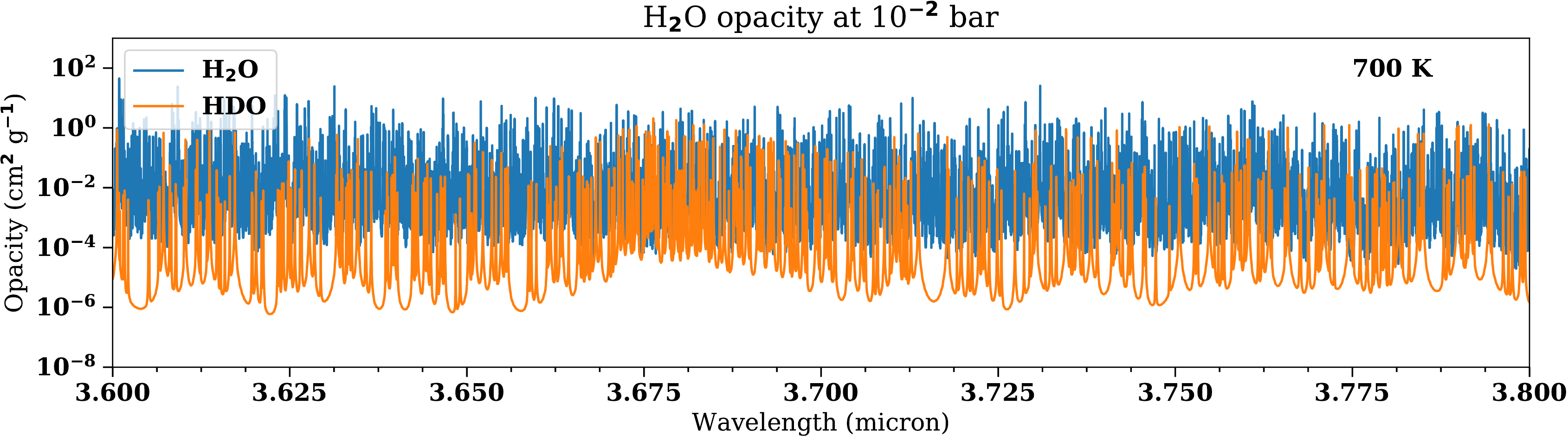}
\includegraphics[width=0.99\textwidth]{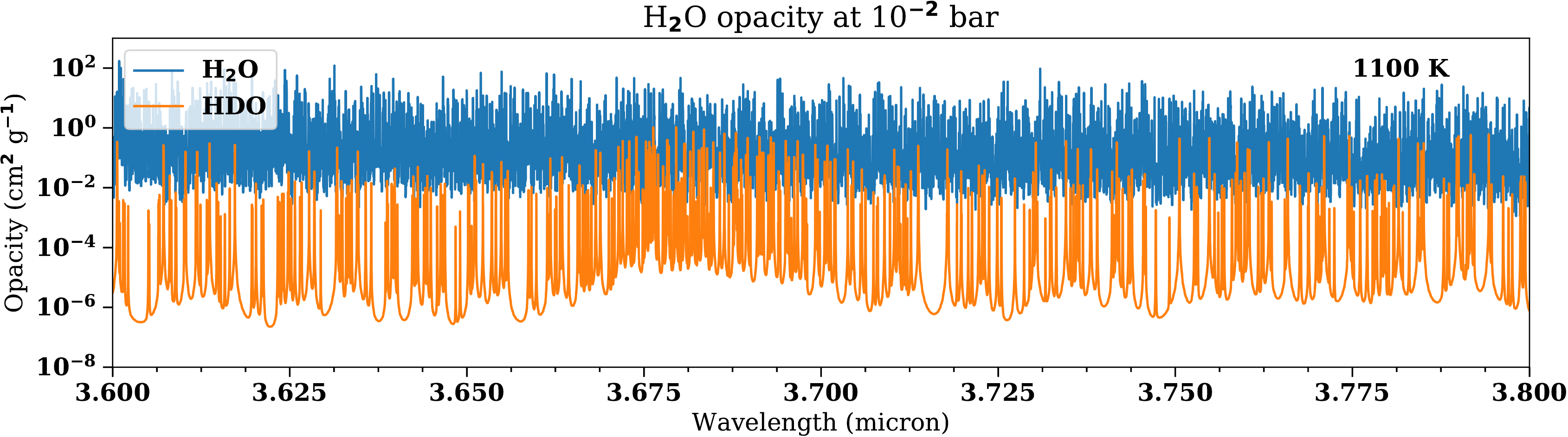}

\caption{Opacities of H$_2$O (blue) and HDO (orange) at $T=300$~K ({\it upper panel}), $T=700$~K ({\it middle panel}) and $T=1100$~K ({\it lower panel}), at \rch{$P=10^{-2}$~bar, which is a representative pressure value probed by water and HDO in the calculations considered here.} The opacities have been scaled such that H$_2$O constitutes 99.7~\% and HDO $4\times 10^{-5}$ of all H$_2$O molecules. The higher the temperature, the more shielding of HDO occurs by H$_2$O.}
\label{fig:h2o_38_micron}

\end{figure*}

For reference, Figure \ref{fig:wlen_study} also shows the transmission of the Earth atmosphere (light grey) and the logarithm of the opacity of the carbon monoxide molecule as function of wavelength (light red). As expected, because the isotope-mass differences are small, the main and secondary isotopologue can only be detected where the CO opacity is high, that is at $\sim$2.4 microns and around $\sim$4.7 microns. \rch{Our simulations are shown for fixed exposure times (28 seconds per spectrum), and include the stellar and sky photon noise variations as a function of wavelength. With this we find that the 2.4-micron region is more efficient for the $^{13}$C$^{16}$O detection, with an expected SNR of 5 for the assumptions given above. In the case studied here, sky background and stellar photon noise are of equal importance at 4.7 microns, leading to an SNR of $\sim$4.5 for the $^{13}$C$^{16}$O detection.} 

Figure \ref{fig:wlen_study} also shows the literature SNR of the CO detection in the 2.3 micron region, as reported by \citet{brogidekok2014}. Their SNR value (5.8) was scaled up to 9 to account for the fact that our wavelength bins are broader. Their detection SNR is broadly consistent at roughly 80\% of the prediction presented here, resulting from data with overall a somewhat lower signal to noise. Hence we expect $^{13}$C$^{16}$O to be detectable with \emph{CRIRES+} on the VLT.

\section{Detecting HDO in atmospheres of self-luminous \rch{and irradiated} planets}
\label{sect:hdo_seld_lum}

In this section, we \rch{mainly} focus on the detectability of HDO in the thermal spectra of young, self-luminous, gas-dominated planets. In the light of planet formation and evolution, a planet's D/H value is arguably the most interesting isotope ratio to study.  Both atmospheric evaporation and icy planetesimal accretion can have a noticeable impact, both tending to increase the atmospheric D/H. In contrast, substellar objects more massive than 13~$\mj$ can burn their deuterium, regardless of their formation pathway \citep{saumon1996,chabrier2000,burrows2001,baraffechabrier2003,molliere2012,bodenheimer2013}, and  observations of deuterium in these objects  has been suggested as a test for whether their mass is above or below the deuterium burning threshold \citep{bejorosorio1999,pavlenkoharris2008}. 

The main wavelength region of interest for HDO measurements is around 3.7 microns, where at low temperatures ($T\lesssim 600$~K), and for galactic abundances (HDO/H$_2$O $\approx$ 4 $\times 10^{-5}$) the HDO opacity protrudes through a minimum of the opacity of the main isotopologue of H$_2$O. For example, this  wavelength region  has been used to study the HDO abundances on Mars \citep[see, e.g.,][]{villanuevamumma2015}. The detectability of HDO at a given D/H is expected to strongly depend on atmospheric temperature. While it is obvious that thermal emission from cool planets is more difficult to detect than that from warm planets, the 3.7 micron region is, in contrast, relatively clean from H$_2$O opacity at low temperature, but largely blanketed by it in hotter atmospheres. This is shown in Figure \ref{fig:h2o_38_micron}, where the water opacities in the 3.7 micron region are plotted relative to those of HDO for different temperatures.\footnote{We use \emph{HITEMP} water opacities, for which the secondary isotopologue lines are taken from the \emph{HITRAN} line list. \emph{HITRAN} is known to be incomplete at high temperatures, but also the high-temperature \emph{Exomol} line lists both for H$_2$O \citep{barbertennyson2006} and HDO \citep{voronintennyson2010} exhibit this behaviour when inspected with the Exomol cross-section service \citep{hillyurchenko2013}.} We therefore expect that HDO will be best detected in relatively cool, directly imaged planets, such that the angular separation from their host star leads to a much decreased stellar flux and noise in the planet spectrum.
\rch{We note that we also study the cases of two irradiated planets, within the framework of this section. This is only an approximation, because the planetary spectra used in the analysis below assumed self-luminous planets.}

Unfortunately, HDO measurements may be hampered by blanketing by CH$_4$ absorption in the same wavelength range. However, the latter may be quenched, meaning that methane-poor gas is mixed up from deeper, hotter regions in the atmosphere. This particular case is therefore also studied below (see Section \ref{sect:hdo_detect_ch4_quench}). We investigate the detectability of CH$_3$D in Section \ref{sect:ch3d_self_lum}.

\subsection{Synthetic observations}
\label{sect:synth_obs_hdo}

The atmospheric structures were calculated using the \emph{petitCODE} as introduced in Section \ref{sect:atmo_struct}, for which we assumed self-luminous planets, with a surface gravity of ${\rm log}_{10}(g)=3.5$ (cgs), a solar composition (${\rm [Fe/H]} = 0$), and \rch{equilibrium} temperatures varying from 300 to 1500~K in $\Delta T=100$~K steps. Clouds were neglected in the calculations. These  are models modified from the atmosphere grid calculated for \citet{samlandmolliere2017}, where \emph{HITRAN} opacities are used for NH$_3$ and PH$_3$ in the structure calculations. Here we use their \emph{Exomol} counterparts for the high-resolution calculations.

Since non-equilibrium chemistry can quench the CO, CH$_4$ and H$_2$O abundances in lower-temperature planets, by mixing up CO-rich and CH$_4$-poor material from high-temperature, high-pressure regions of the planets \citep[see, e.g.,][]{zahnlemarley2014}, we consider models both with equilibrium abundances, as well as models where CH$_4$ (and CO$_2$) has been excluded in the chemical equilibrium calculations, constituting an extreme quenching scenario (see Section \ref{sect:hdo_detect_ch4_quench}).  The exclusion of CO$_2$ was necessary because equilibrium chemistry  would otherwise lock up oxygen in this molecule, which should stay in H$_2$O in the real quenching case. We also found that the carbon in CO is preferentially moving into C$_2$H$_2$ at low pressures and temperatures ($T\lesssim 140 \ {\rm K}$), but CO does not have any features in the 3.7 micron region, so this effect was neglected here.

Subsequently, high-resolution spectra for the planets were calculated as described in Section \ref{sect:high_res_code}, taking into account the opacities of H$_2$O, CO, H$_2$S, NH$_3$, PH$_3$, CH$_4$ and CO$_2$, as well as H$_2$--H$_2$ and H$_2$--He CIA. Nominally, the HDO/H$_2$O ratio was assumed to be twice the cosmic D/H value of $2\times 10^{-5}$. The factor two arises from combinatorics, that is the fact that every water molecule has two locations where D may be placed, instead of H, when forming HDO instead of forming H$_2$O (this will be a factor four for CH$_3$D). The wavelength region considered here was from 3.6 to 3.8 microns. As before, we assumed a resolution of $10^5$, and three wavelength steps per resolution element.

\subsection{Analysis of the synthetic observations}
\label{sect:synth_obs_red_hdo}

The analysis of the synthetic observations is similar to that for CO described in Section \ref{sect:data_red}. The main difference is that the planet does not exhibit any measurable change in its radial velocity offset during the observations (assuming a long $>$year orbital period). Therefore, it is not the change in Doppler shift that is used to separate the planet spectral features from that of the star and the Earth atmosphere. Instead, both the planet and star are observed simultaneously, but angularly separated, allowing the stellar spectrum to be used for removing the stellar and telluric contributions from the planet spectrum \citep{snellendekok2015}.

Template planet spectra $F_{\rm P}(\lambda)$ and $F_{\rm P - HDO}(\lambda)$ are created, which are identical except that the latter has its HDO opacity removed. Comparing to Section \ref{sect:data_red}, $F_{\rm P}(\lambda)$ corresponds to data set A (without tellurics and noise) and $F_{\rm P}(\lambda) - F_{\rm P - HDO}(\lambda)$ to data set B. 
In principle, subsequent analysis would involve the addition of noise and telluric absorption, followed by reduction steps similar as for the hot Jupiter case described in Section \ref{sect:synth_obs}. However, since this procedure must be performed many times, we derived, tested, and used the following equation to approximate the statistical detection level of HDO for an observation with a certain SNR per wavelength step:
\begin{align}
\nonumber S/N & = \frac{1}{\sigma} \left\{  \sum_{i=1}^{N_{\rm \lambda}}    \left[ F_{\rm P}(\lambda_i) - F_{\rm P - HDO}(\lambda_i) \right]^2 \right\}^{1/2} \\
 & = \frac{1}{\left<F_{\rm P}\right>}\left(S/N\right)_{\rm pix}\left\{\sum_{i=1}^{N_{\rm \lambda}}\left[F_{{\rm P}}(\lambda_i)-F_{{\rm P-HDO}}(\lambda_i)\right]^2\right\}^{1/2} \ ,
\label{equ:SNR_predict}
\end{align}

where $\sigma$ is the error in the spectrum per wavelength step, $N_\lambda$ is the number of spectral points, $(S/N)_{\rm pix}$ is the average SNR per wavelength step, and $\left<F_{\rm P}\right>$ is the average flux per wavelength step in the targeted spectrum. The derivation of Equation \ref{equ:SNR_predict} is given in Appendix \ref{app:cross_corr_SNR_lines_var_strength}.  This approximative formula predicts the SNR of the HDO detection when cross-correlating data set B with the noiseless B model (corresponding to the red solid line in the right panel of Figure \ref{fig:test_CCFs}).

This formula was tested by comparing it to the full synthetic analysis (i.e. adding tellurics and noise, reducing the data, cross-correlating with an HDO template), leading to a good agreement, see Appendix \ref{sect:first_HDO_test}. The reader should note that due to the complexity of the HDO spectrum, and methane absorption that causes a quasi-continuum depending on the atmospheric temperature, even simpler SNR estimates as mentioned for CO in Section \ref{ref:cross_corr_CO} (SNR scaling with $N_{\rm lines}^{1/2}$, where $N_{\rm lines}$ is the number of lines) are not adequate.

\rch{We also want to point out that Equation \ref{equ:SNR_predict} predicts the SNR accurately, when compared to the full analysis in Appendix \ref{sect:first_HDO_test}, even though the effects of telluric absorption are not included in Equation \ref{equ:SNR_predict}. The full analysis in Appendix \ref{sect:first_HDO_test} includes the effect of telluric absorption, the spectrum of which is actually dominated by telluric HDO. The telluric HDO does not absorb the exoplanet HDO signal because we assumed a Doppler shift of 30 km/s between the exoplanet and Earth, see Appendix \ref{sect:first_HDO_test}, which shifts the HDO lines enough to fall outside the telluric HDO lines.}
\subsection{Required spectral SNR to detect HDO}
\label{sect:SNR_requ_hdo_detect}

\begin{figure*} 
\centering
\includegraphics[width=1.0\textwidth]{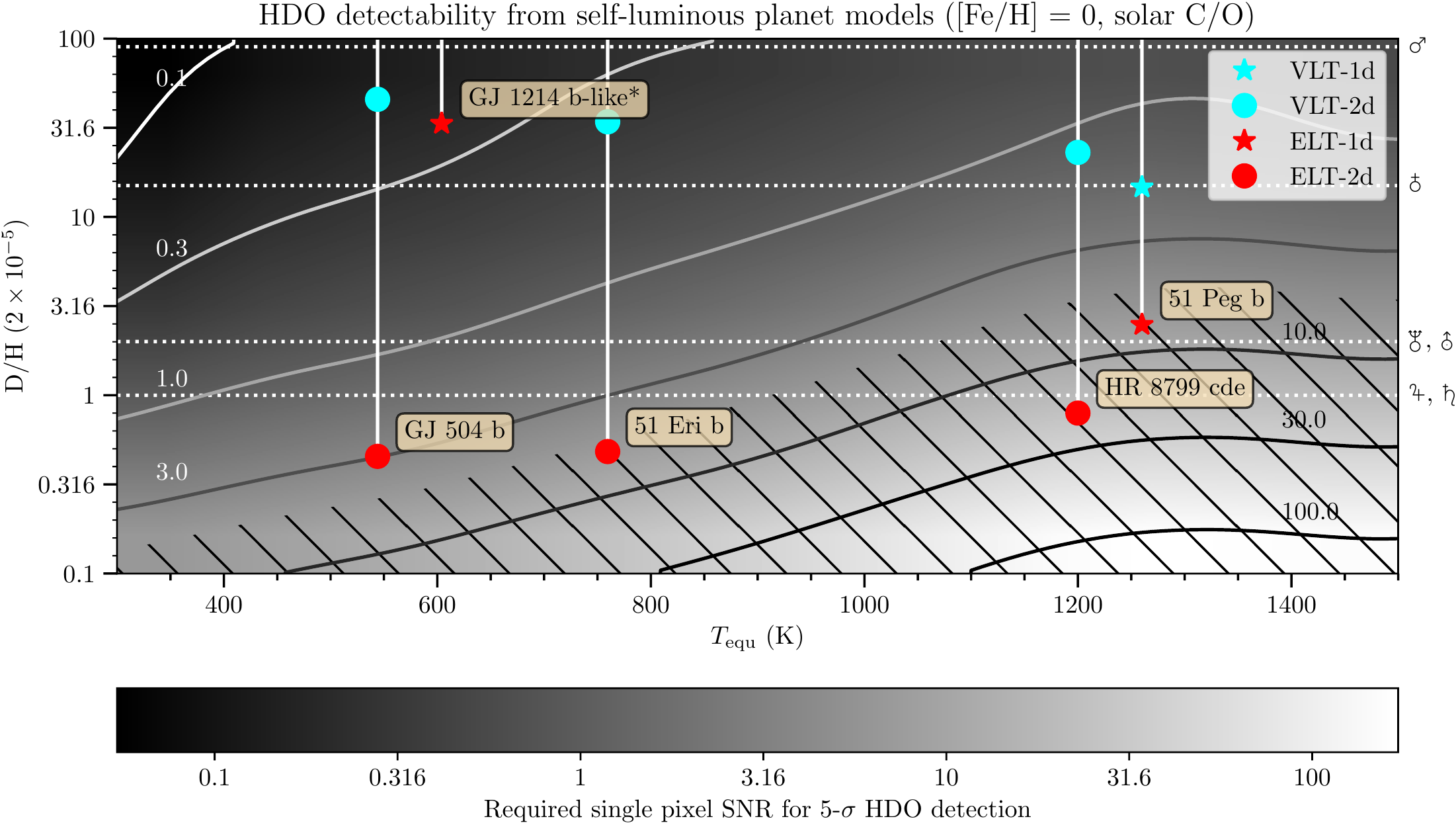}
\caption{Contour map showing the required SNR of the planetary spectrum per pixel, as function of planetary \rch{equilibrium} temperature and D/H value, for an HDO detection at an SNR of five.  
The considered wavelength region is 3.6 to 3.8 microns. The D/H values of the Solar System planets are indicated by the horizontal white dotted lines. The colored symbols indicate the lowest detectable D/H values for various exoplanets. Cyan and red symbols indicate limits for \emph{CRIRES+}@VLT and \emph{METIS}@ELT, respectively, assuming a single night of observation (10 h). Filled circles stand for those targets which can be angularly separated from their host star, assuming a stellar flux reduction at the planet position by a factor 100 and 1000 for the VLT and ELT, respectively.  Star-symbols denote planets that cannot be spatially resolved, for example hot Jupiters, hence no flux suppression is possible. $^*$\emph{GJ 1214\,b-like} planet is a hypothetical non-transiting twin of GJ 1214\,b at half the distance from Earth. The hatched area indicates the region where the required SNR per pixel is larger than 5, implying very weak HDO lines, which may be difficult to recover accurately with planet atmospheric modeling.}
\label{fig:hdo_detect_eq_chem}
\end{figure*}

In Figure \ref{fig:hdo_detect_eq_chem} we show the required SNR per pixel of a planet spectrum for detecting HDO at a SNR of 5, in the 3.6 to 3.8~$\mu$m region, as a function of planetary $T_{\rm equ}$ and D/H value. These were calculated using Equation \ref{equ:SNR_predict}.

The required SNR per pixel increases as function of temperature, for example for the cosmic D/H value from $\sim$1 at 400 K to $\sim$15 at 1200 K, due to the increased shielding by H$_2$O absorption at higher temperatures. The 3.7 $\mu$m region is relatively clear of water absorption at low temperatures. The reader should note, however, that it is generally more difficult to reach a certain SNR level for a cooler planet than for a warmer planet. 
The required SNR per pixel flattens out at higher temperatures due to the expected decrease in methane abundance. Obviously, HDO is easier to detect if the atmospheric D/H value is higher. At 400 K, only an SNR per pixel of 0.1 is needed if the D/H value is 100 times the cosmic value (e.g. that of Mars). 

It is instructive to compare the results presented in Figure \ref{fig:hdo_detect_eq_chem} with the expected SNR limits of known exoplanets achievable with the current 10m-class telescopes and the future Extremely Large Telescopes (ELTs). For this we concentrate on the \emph{CRIRES+} instrument \citep{follertdorn2014} on ESO's VLT (cyan symbols) and \emph{METIS} \citep{brandlfeldt2014} (red symbols) on the European ELT, and a few prototypical exoplanets indicated in Figure \ref{fig:hdo_detect_eq_chem}, assuming a single night (10 hr) of observations. The planetary parameters used for this study are given in Table \ref{tab:planet_params_hdo_ch3d}.

For \emph{CRIRES+} on the VLT, we assumed an instrument resolution of $R=100,000$, three pixels per resolution element, a mirror surface area of 52~m$^2$, and a total telescope+instrument throughput of 0.15. For \emph{METIS} on the ELT, identical specifications were assumed, but the mirror surface area was changed to 976~m$^2$. The planetary flux was estimated by interpolating our synthetic, self-luminous exoplanet models to the planets' published equilibrium temperature, and subsequently using the planet radius and distance to the Solar System to scale the flux accordingly. The effect of the planetary ${\rm log}(g)$, composition, and cloudiness are hence neglected in the SNR estimates presented here. The stellar flux was obtained in the same way, using \emph{PHOENIX} models \citep{hauschildtbaron1999b} for the host star spectra, as described in \citet{mollierevanboekel2015}. Finally, the expected SNR per pixel for one night of observations was obtained by computing the mean number of photons per instrument pixel of planet, star, \rch{and the sky emission}, and then calculating
\beq
(S/N)_{\rm pix} = \frac{N_{\rm P}}{\left(N_{\rm P}+N_{*}/f+\bar{N}_{\rm sky}\right)^{1/2}},
\label{equ:SNR_pix_estimate}
\eeq
where $N_{\rm P}$, $N_{*}$, \rch{and $\bar{N}_{\rm sky}$} are the number of photons per pixel of planet, star, \rch{and the wavelength median of the sky background}, respectively. The factor $f$ denotes the amount of starlight reduction at the planet position. In the case that planet and star are not angularly separated, denoted as VLT-1d and ELT-1d, there is no starlight reduction and $f=1$. For directly imaged planets, denoted as VLT-2d and ELT-2d in Figure \ref{fig:hdo_detect_eq_chem}, $f$ is assumed to be 100 and 1000 for VLT-2d and ELT-2d observations, respectively, using slit spectroscopy \rch{\citep[e.g.][]{snellenbrandl2015}}, or the integral field unit in the case of \emph{METIS} \citep[e.g.][]{snellendekok2015}. \rch{As discussed in the end of Section \ref{sect:synth_obs_red_hdo} above, the telluric absorption, and hence emission, is dominated by telluric HDO, which is Doppler-shifted with respect to the planet signal. We are hence interested in the average sky emission between the telluric HDO lines, which we estimated by calculating the median sky emission in the 3.6 to 3.8 micron region.}

\rch{While the sky background is negligible for the VLT-1d and ELT-1d cases, where $f=1$, we find that the photon noise of star and planet can often be comparable to the sky background noise for the VLT-2d and ELT-2d cases, which have $f=100$ and 1000, respectively.}

\begin{figure*} 

\includegraphics[width=1.0\textwidth]{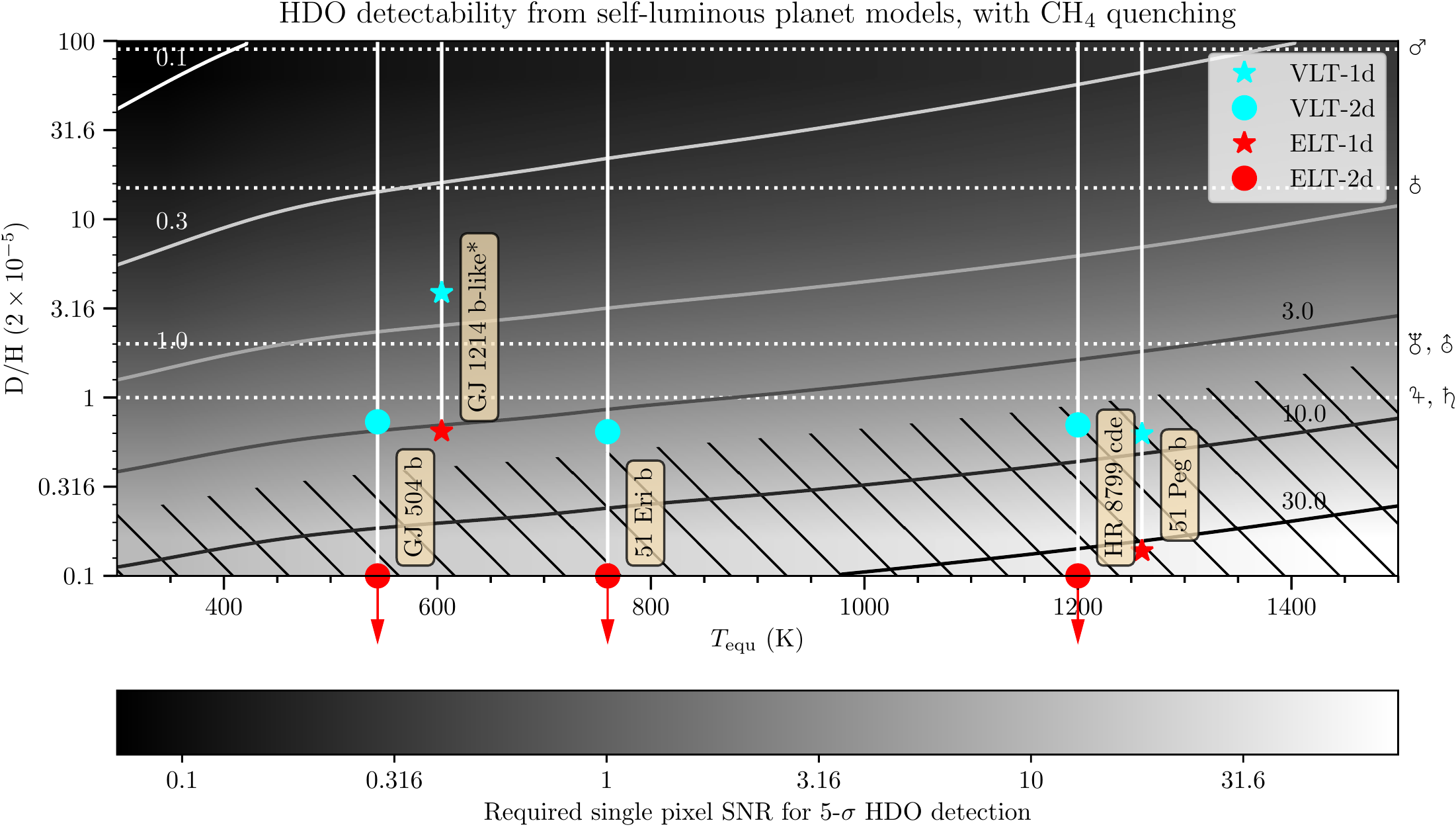}
\caption{Same as Figure \ref{fig:hdo_detect_eq_chem}, but neglecting both CH$_4$ (and CO$_2$) in the chemistry and for the opacities, in order to mimic an atmosphere where CH$_4$, CO, and H$_2$O are quenched at high pressures (i.e. high temperatures).}
\label{fig:hdo_detect_eq_chem_no_ch4}
\end{figure*}

\subsubsection*{Individual planets:}
\subsubsection*{GJ~504b, 51~Eri~b, HR~8799~cde}
For the directly imaged planets GJ~504b, 51~Eri~b, and HR~8799~cde, assuming equilibrium chemistry, we expect that galactic D/H values will be out of reach for 10m class telescopes. With the ELT, galactic D/H values are all in reach within a single night. 

\subsubsection*{Super Earths: GJ~1214b-like planets}
GJ~1214\,b (and other planets like it) is an interesting target since it is of low mass \citep[6.5~${\rm M}_\oplus$][]{charbonneauberta2009}), significantly irradiated ($T_{\rm{equ}} = 604$~K, see Table \ref{tab:planet_params_hdo_ch3d}), and potentially highly enriched in icy planetesimals, and therefore could have a high D/H value. It is not expected that recently formed, self-luminous planets of this mass would ever exhibit such temperature, due to the low amount of formation heat retained  \citep{lindermordasini2018}. While GJ~1214\,b itself is too faint, similar non-transiting systems should be found at smaller distances \rch{from the Solar System}. With a transit probability of $\sim$7\%, the nearest non-transiting GJ~1214\,b-like system is expected to be found at approximatly half the distance, with a host star \rch{1.5} magnitudes brighter, which we used for our simulations.  

Using \emph{METIS} on the ELT, we expect that atmospheric D/H values $\geq$30 times the galactic mean value may allow for the detection of HDO in a single night. Current theories point to GJ~1214~b being strongly enriched in metals \citep[by a factor 100 to 1000 w.r.t. solar, see][]{morleyfortney2013,morleyfortney2015,mollierevanboekel2017}, a large D/H value may well be expected for this type of planet (see the discussion in Section \ref{sect:intro}), although probably not as high as 30. We assume solar abundances in the spectral models used here, so only D/H varies. Moreover, the spectral models used here were for self-luminous planets, while the GJ~1214~b-twin planet is irradiated. We will revisit this planet in the methane quenching and \ce{CH3D} detection cases in section \ref{sect:hdo_detect_ch4_quench} and \ref{sect:ch3d_self_lum}.

\subsubsection*{51~Peg~b}
Similar to the GJ~1214b-like case studied above, we use our self-luminous atmospheric grid to study the detectability of HDO in the atmosphere of the hot Jupiter 51~Peg~b. While this planet is not a self-luminous planet, our analysis gives a first estimate of the single-pixel SNR to be expected for hot Jupiters. Here we predict that one night of observations with VLT \emph{CRIRES+} will allow to detect HDO if the atmospheric D/H value is 15 times the galactic value or higher, while ELT \emph{METIS} would allow detecting HDO down to D/H values of twice the galactic value. For a gas giant like 51~Peg~b one would expect a D/H value similar to the galactic value (see Section \ref{sect:intro}), which could be reached in four nights with ELT \emph{METIS}, thus remaining a challenge.

It is important to note that, for the hotter planets, large SNRs per pixel are required to detect HDO. For HR~8799~cde and 51~Peg~b, the required SNR per pixel is $>$10 for the  detection of HDO at galactic D/H values, implying intrinsicly very weak HDO lines. This requires very accurate and complete planet atmospheric modeling, since all weak planet lines need to be accounted for at this level.  This is something which has to be investigated in future studies. The region where the required SNR per pixel is larger than 5 is hence indicated by the hatched area in Figure \ref{fig:hdo_detect_eq_chem}. \rch{If trying to avoid this hatched region, HDO at galactic abundances may be detectable in planets out to equilibrium temperatures of 900~K}.

\subsection{HDO detection with methane quenching}
\label{sect:hdo_detect_ch4_quench}

The results for the methane-depleted calculations (see Section \ref{sect:synth_obs_hdo}) are shown in Figure \ref{fig:hdo_detect_eq_chem_no_ch4}. As expected, the required single-pixel SNR to detect HDO decreases substantially: HDO at galactic abundances can now be detected in a single night of VLT \emph{CRIRES+} observations in all example planets, except for the GJ~1214~b-like case, which would require ELT observations. But we note that a higher D/H value is likely to be expected for such a planet, as discussed in Section \ref{sect:SNR_requ_hdo_detect}.
 
The required SNR per pixel for these cases is lower than 5 out to 1200~K, implying significantly stronger HDO lines than in the default non-quenching case. The gain in sensitivity is due to the fact that methane is not blanketing the HDO lines, but also because 
we effectively peer into the deeper, warmer and therefore brighter regions of the planet atmosphere. 

Moreover, the required SNR in the quenching case is a monotonously increasing function of $T_{\rm equ}$ for all D/H values. The plateau seen in the non-quenching case is not present and therefore evidently caused by the decrease in methane absorption with temperature.

\section{Detecting CH$_3$D in self-luminous  \rch{and irradiated}  planets}
\label{sect:ch3d_self_lum}
\begin{figure*}[t!]
\centering

\includegraphics[width=0.99\textwidth]{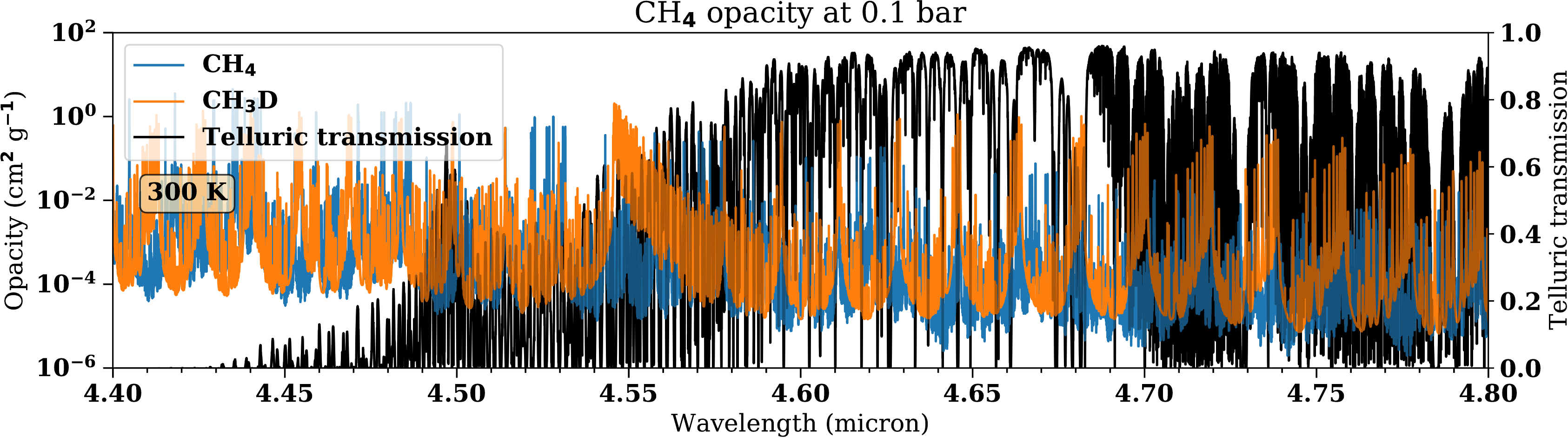}
\includegraphics[width=0.99\textwidth]{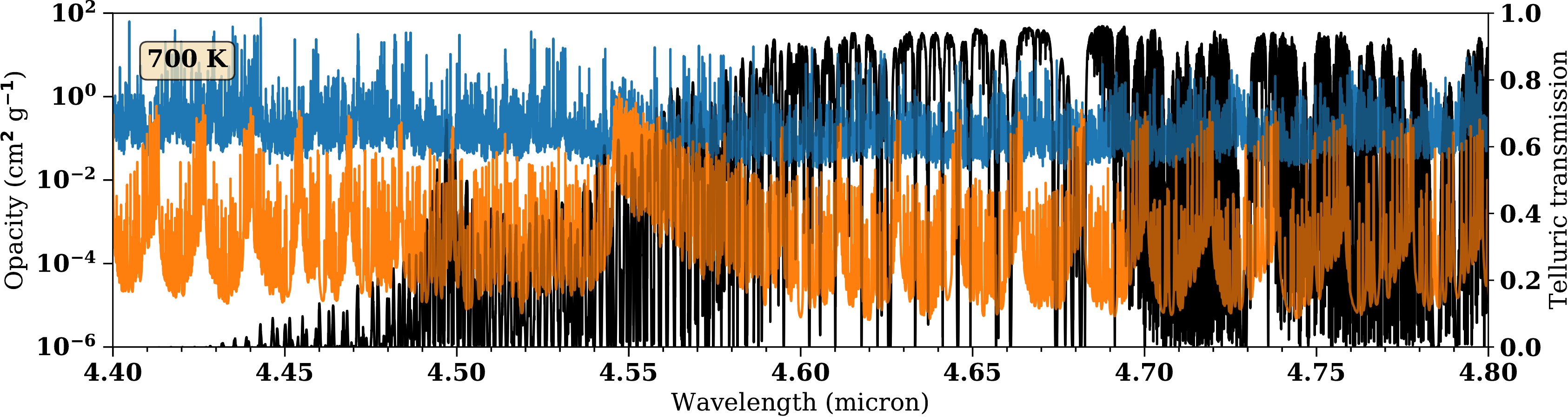}
\caption{Opacities of CH$_4$ (blue) and CH$_3$D (orange), shown at $T=300$~K ({\it upper panel}) and $T=700$~K ({\it lower panel}), at \rch{$P=0.1$~bar, which is a representative pressure value probed by methane and CH$_3$D in the calculations considered here.} The opacities have been scaled such that CH$_4$ and CH$_3$D constitute 98.8~\% and $8\times 10^{-5}$ of all CH$_4$ isotopologues, respectively. The telluric transmission is shown in black.}
\label{fig:ch4_opa}

\end{figure*}

In this section, we study the detectability of the methane isotopologue CH$_3$D, which has been used to infer the D/H value in Uranus, Neptune, Saturn, and Jupiter, and in the atmosphere of Saturn's moon Titan \citep[see, e.g.,][]{debergh1995,owenencrenaz2003}.

Here we focus on the rovibrational CH$_3$D band centered around 4.6~microns. This band has the advantage that both the expected planet CH$_4$ and H$_2$O opacity are comparatively low, making this wavelength range appear ideal for the detection of CH$_3$D. Consequently, this wavelength region has been recently advertised for detecting CH$_3$D in WISE~0855 \citep{skemermorley2016b,morleyskemer2018,morleyskemerb2018}. \rch{One complication of the 4.6 micron region is that the telluric background emission will become the dominating noise source, especially for self-luminous planets, where the planet can be angularly separated from the star.}

Figure \ref{fig:ch4_opa} shows the opacities of CH$_3$D and CH$_4$ at temperatures of 300 and 700~K, and the telluric transmission in the 4.6~micron region.
Similar to the behaviour seen for HDO and H$_2$O (see Figure \ref{fig:h2o_38_micron} and the discussion in Section \ref{sect:hdo_seld_lum}), CH$_3$D is weaker with respect to CH$_4$ at higher temperatures, because the lines in the opacity minimum of CH$_4$ are stronger at high temperature and blanket the CH$_3$D opacity. We use the \emph{HITRAN} line list for CH$_3$D, and \emph{Exomol} for CH$_4$. As the telluric absorption \rch{and emission} is quite strong shortward of 4.6~microns, our analysis concentrates on the range between 4.6 and 4.8~microns. Future analyses may include the relatively strong CH$_3$D feature at 4.55~microns, requiring highly accurate telluric corrections.

In terms of opacity and chemical abundances, we expect CH$_3$D to be most easily detected in cool exoplanets, where CH$_3$D is sufficiently strong compared to CH$_4$, and equilibrium chemistry predicts high methane abundances for H$_2$/He dominated atmospheres. Hotter planets are expected to have a lower methane abundance both due to chemical equilibrium effects and methane quenching \citep[see, e.g.][]{zahnlemarley2014}, as discussed earlier.

\subsection{Synthetic observations and analysis}
\label{sect:synth_obs_ch3d}

The same atmospheric models of self-luminous gas giant planets were used as for the HDO study, focussing on the wavelength range from 4.6 to 4.8 microns, and assuming a \rch{nominal} CH$_3$D abundance of $8\times10^{-5}$, relative to CH$_4$.\footnote{The 4-fold increase when compared to the galactic mean value ($2\times10^{-5}$) is again caused by combinatoric effects.} All nominal isotopologue abundances used in this paper can be found in Table \ref{tab:nom_iso_abund}.

\label{sect:synth_obs_red_ch3d}

We use the same analysis technique as described in Section \ref{sect:synth_obs_red_hdo}. However, Equation \ref{equ:SNR_predict} was modified, since this wavelength region is more strongly affected by telluric absorption \rch{and emission}. The derivation follows the same principles outlined in Appendix \ref{app:cross_corr_SNR_lines_var_strength}, but accounts for the flux attenuation by tellurics, leading to scaling of the flux with $\mathcal{T}$. \rch{For irradiated planets around bright hosts, where the stellar photon noise still dominates, the noise is scaled with $\mathcal{T}^{1/2}$, where $\mathcal{T}$ is the telluric transmission. For self-luminous planets the stellar photon noise is replaced by the photon noise of the sky background:}

\begin{align}
S &=  \sum_{i=1}^{N_{\rm \lambda}}    \mathcal{T}(\lambda_i)\left[ F_{\rm P}(\lambda_i) - F_{\rm P - \ce{CH3D}}(\lambda_i) \right]^2, \label{equ:SNR_predict_ch3d1} \\
N_{\rm irrad} &= \left\{\sum_{i=1}^{N_{\rm \lambda}}    \left[ F_{\rm P}(\lambda_i) - F_{\rm P - \ce{CH3D}}(\lambda_i) \right]^2\sigma_{\rm telluric}^2(\lambda_i)\right\}^{1/2}, \label{equ:SNR_predict_ch3d2_bright}\\
N_{\rm sky} &= \left\{\sum_{i=1}^{N_{\rm \lambda}}    \left[ F_{\rm P}(\lambda_i) - F_{\rm P - \ce{CH3D}}(\lambda_i) \right]^2\sigma_{\rm sky}^2(\lambda_i)\right\}^{1/2},
\label{equ:SNR_predict_ch3d2_bg}
\end{align}
\rch{where the effective single-pixel noise for cases dominated by stellar photons is defined by $\sigma_{\rm telluric} = \mathcal{T}^{1/2}\sigma_{\rm clear}$, with $\sigma_{\rm clear}$ being the photon noise when neglecting the impact of the  Earth's atmosphere. In the background dominated case, the wavelength-dependent form of $\sigma_{\rm sky}$ is again obtained from \emph{SkyCalc}.} The SNR of the CH$_3$D detection is subsequently given by the ratio of equations \ref{equ:SNR_predict_ch3d1} and \ref{equ:SNR_predict_ch3d2_bright}\rch{, or \ref{equ:SNR_predict_ch3d1} and \ref{equ:SNR_predict_ch3d2_bg}, depending on whether stellar or sky background emission dominates the noise budget}.

Analogous to the HDO detection case, this approximation \rch{for a wavelength-dependent noise source} was tested by comparing to the full synthetic analysis (i.e. adding tellurics and noise, reducing the data, cross-correlating with a CH$_3$D template), again leading to good agreement, see Appendix \ref{sect:first_CH3D_test}.

\begin{figure*}[t!]
\centering
\includegraphics[width=1.0\textwidth]{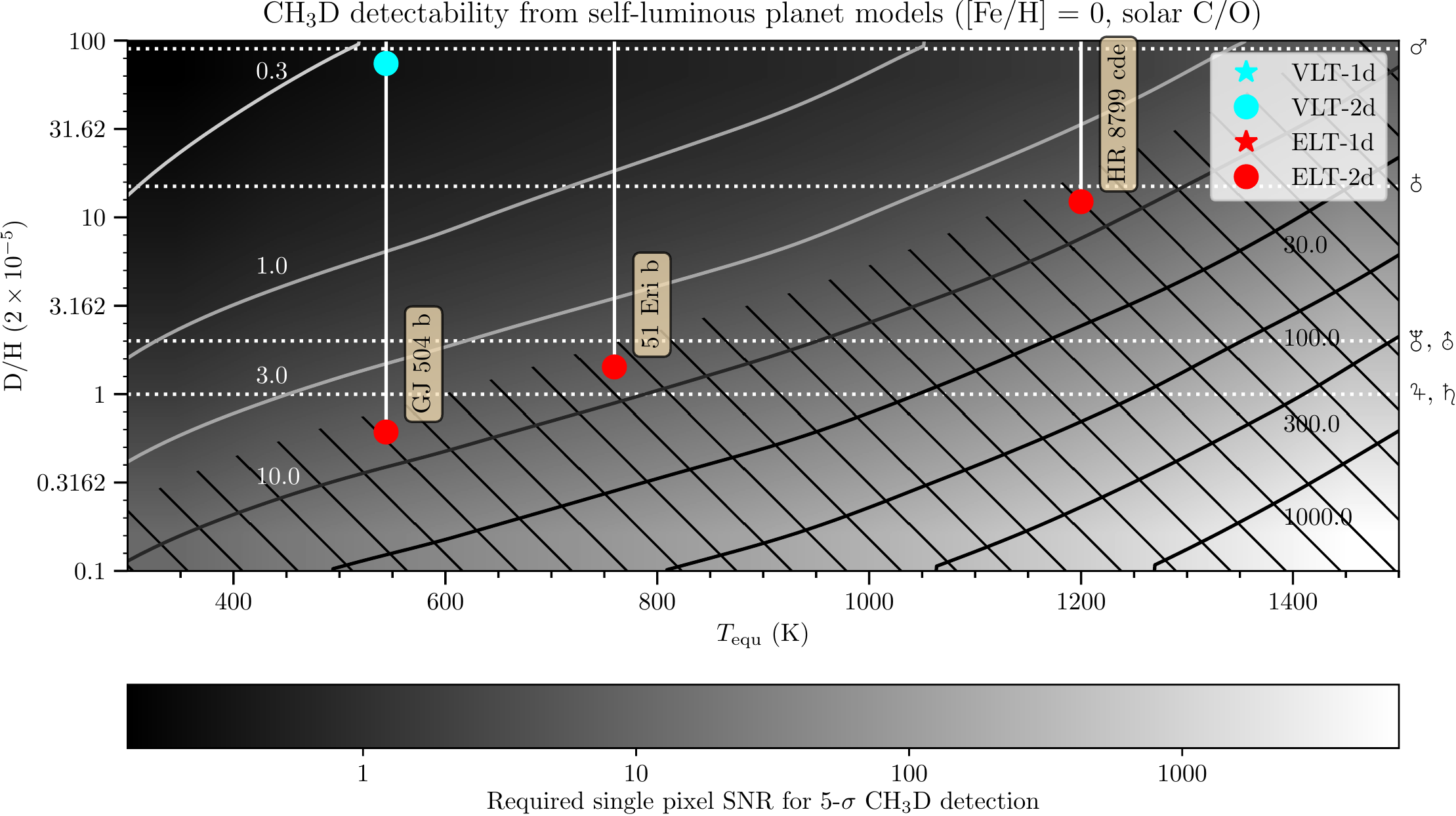}
\caption{Same as Figure \ref{fig:hdo_detect_eq_chem}, but for CH$_3$D. \rch{The single pixel SNR is defined as the mean planet signal per pixel, in the 4.6 to 4.8 micron range, divided by the median sky background noise per pixel, in the same wavelength range.}}
\label{fig:ch3d_detect_eq_chem}
\end{figure*}

\subsection{Required SNR to detect CH$_3$D}
\label{sect:SNR_requ_ch3d_detect}

In Figure \ref{fig:ch3d_detect_eq_chem} we show the required SNR per pixel of planetary spectra, to detect CH$_3$D at an SNR of 5, as a function of D/H and planetary equilibrium temperature, in the 4.6 to 4.8 micron region. Analogous to the HDO study in Section \ref{sect:hdo_seld_lum}, we overplot the expected SNRs for known planetary systems, assuming a single night (10 hr) of observations for the VLT and ELT. \rch{The plot is presented for cases dominated by sky background emission, so only the self-luminous planets are shown here. For producing this plot, the single pixel SNR was defined by dividing the mean planet signal per pixel in the 4.6 to 4.8 micron region by the median sky background noise, in the same wavelength region. Thus, although the underlying predictions account for the full wavelength dependence of the planet and background spectrum, a single scalar SNR value can be assigned to every ($T_{\rm equ}$, D/H) pair.}

\rch{For cooler planets ($T_{\rm equ}\lesssim 600$~K), CH$_3$D may be detectable using \emph{METIS} on the ELT.
Hence, due to the sky background emission, we expect that CH$_3$D is slightly more difficult to detect than HDO. In the case of strong methane quenching (cf. Figure \ref{fig:hdo_detect_eq_chem_no_ch4}), HDO is significantly easier to detect than CH$_3$D, and potentially even VLT-class telescopes could achieve this.}

\rch{We find that the irradiated GJ~1214~b-like planet, when studied with the ELT, as well as the hot Jupiter 51~Peg~b, when studied with the VLT or ELT, should stay in the regime dominated by stellar photon noise. For the GJ~1214~b-twin, a single night of ELT observations could probe D/H values down to 1.5 times the galactic mean value. For the hotter 51~Peg~b, even the ELT will only probe down to D/H values of $\sim$5 in a single night, requiring a mean SNR of 20 per pixel, which is an unlikely precision to be reached from a modeling standpoint alone. Moreover, D/H values of 5 times the galactic mean value are not expected for gas giant planets.}

\rch{We thus find that CH$_3$D is a disfavored isotopologue when trying to infer D/H values in self-luminous planets, when compared to HDO, because the sky background emission in the 4.7~micron region is too strong. From the ground, HDO represents the better choice at all temperatures, regardless of whether methane is quenched or not. From space, and in the absence of methane quenching, the situation is reversed, however, and the associated CH$_3$D detectability has been advertised for medium resolution spectra with the upcoming \emph{JWST} \citep{morleyskemerb2018}. For irradiated nearby super-Earths, such as the GJ~1214b-like planet studied here, a ground-based CH$_3$D detection may be feasible, especially as a large atmospheric metal enrichment inferred for GJ~1214b \citep[][]{morleyfortney2013,morleyfortney2015,mollierevanboekel2017} may lead to increased D/H values.} 

\section{Detecting HDO in Proxima Centauri~b}
\label{sect:proxima_hdo}

\begin{figure*}[t!]
\centering
\includegraphics[width=0.483\textwidth]{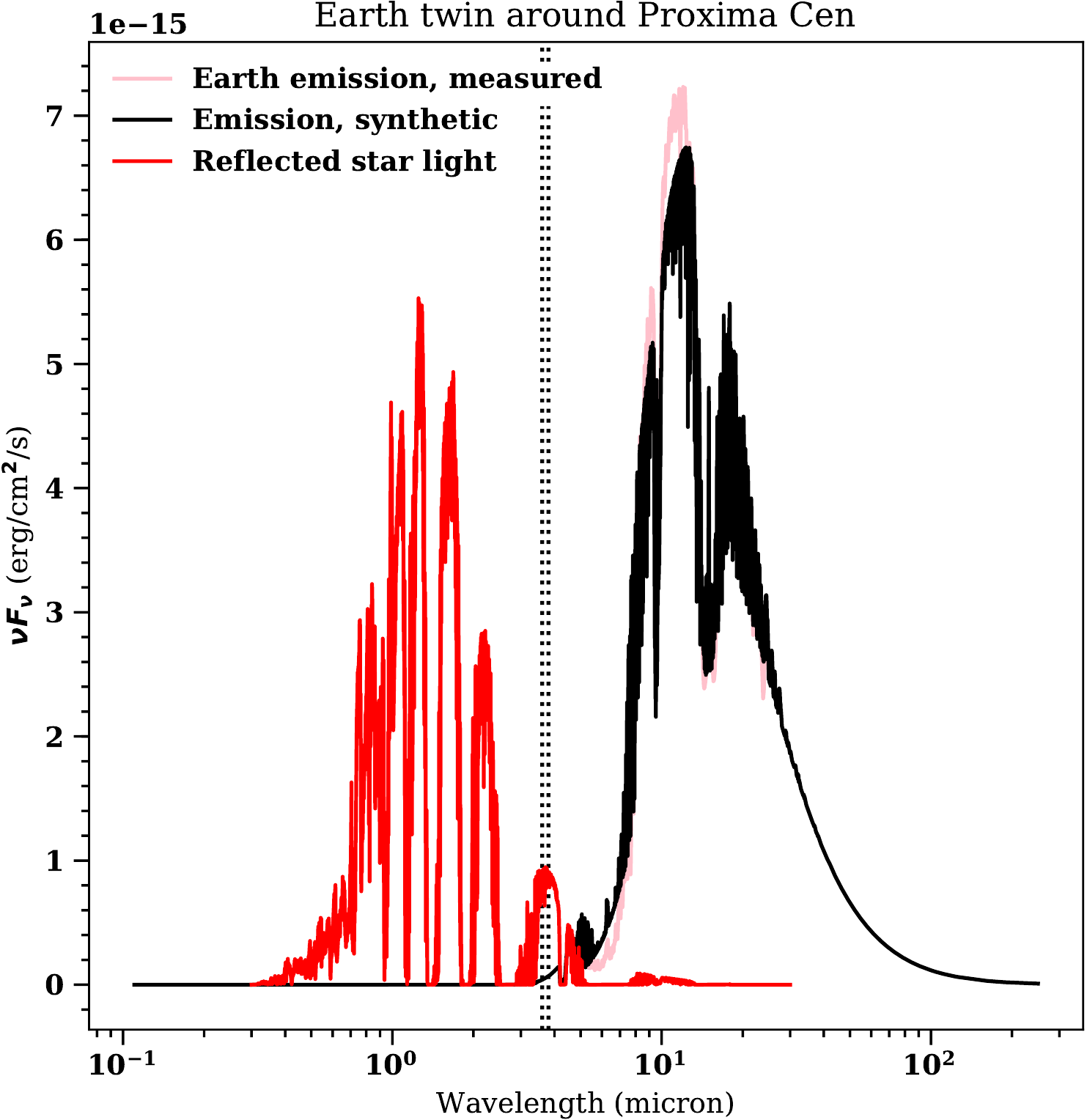}
\includegraphics[width=0.495\textwidth]{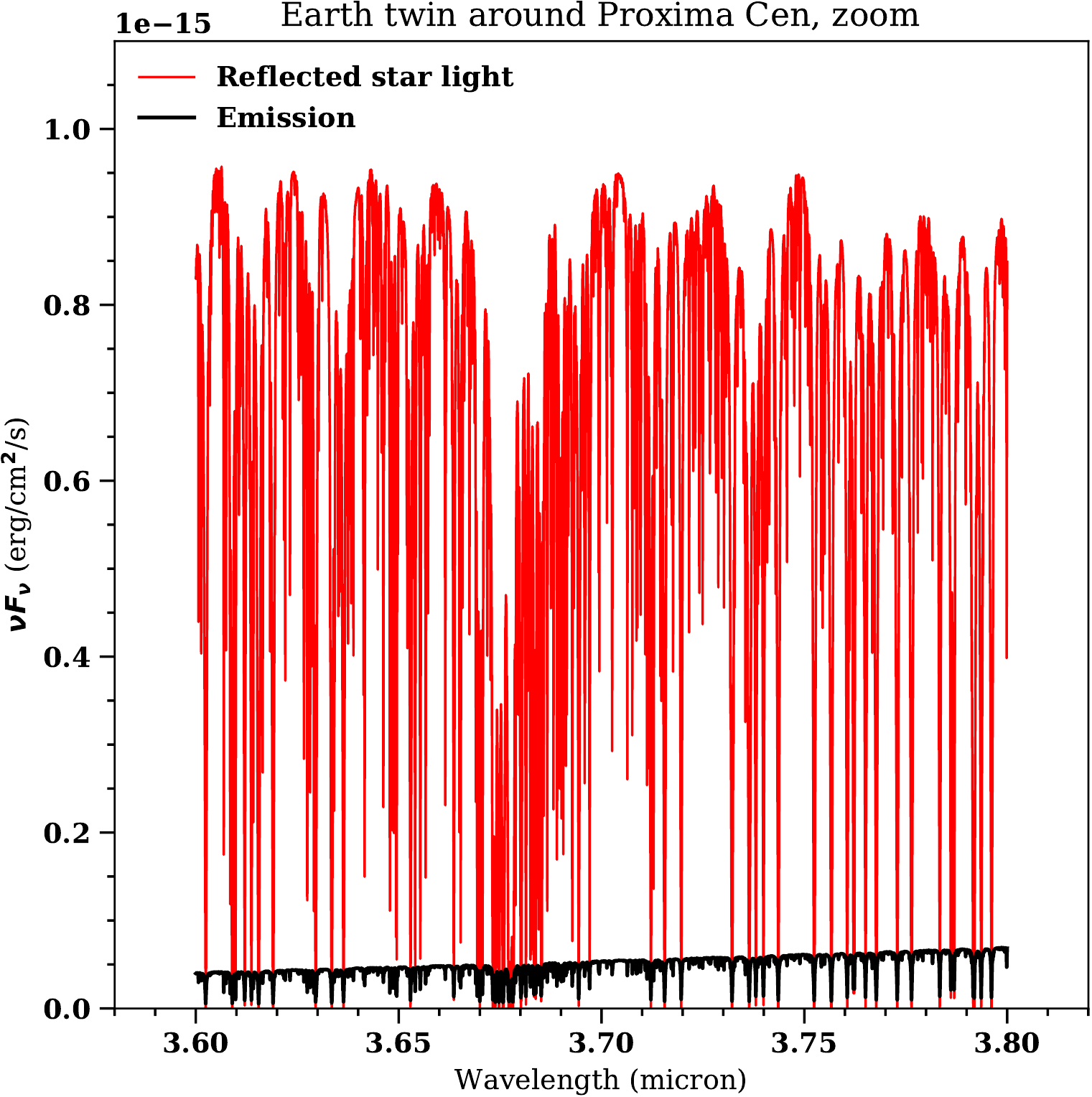}
\caption{Apparent spectrum of Proxima Cen b at a distance of 1.29 pc, assuming it is identical to Earth. {\it Left panel}: The planet synthetic thermal spectrum, assuming the Earth's $P$-$T$ profile and abundances is shown as a black solid line. As a comparison, measurements by the \emph{Nimbus~4} satellite are shown as a pink solid line, which largely overlap. The red solid line shows the reflected light assuming a surface albedo of 30~\%, \rch{including the attenuation of the planetary atmosphere}. The two vertical black dashed lines denote the wavelength region of interest for probing HDO (3.6 to 3.8~$\mu$m). {\it Right panel:} The same but focused on the 
3.6 to 3.8~$\mu$m region, \rch{and showing high-resolution spectra}. At these wavelength, the reflected starlight is $\sim$15 times stronger than the intrinsic thermal emission from the planet.}
\label{fig:HDO_prox}
\end{figure*}

Proxima~Centauri~b is a recently discovered planet in the habitable zone of the Sun's closest stellar neighbor \citep{escudeamado}. With $M_{\rm min} = 1.27 ~M_{\oplus}$, only slightly more massive than the Earth, it has likely a mainly rocky composition with a climate allowing possibly liquid water on its surface \citep{ribasbolmont2016,turbetleconte2016}. In this section, we investigate whether HDO could be detected in the spectrum of Proxima~Cen~b, assuming atmospheric properties identical to that of Earth.

The properties of the planet atmosphere remain yet unknown. However, extreme irradiation conditions and (partial) atmospheric loss likely play or have played an important role on this planet: models estimate stellar wind pressures multiple orders of magnitudes higher than experienced by Earth \citep{garraffodrake2016}. In addition, stellar X-ray and EUV fluxes could have caused loss of water by evaporating Proxima~Cen~b's atmosphere, especially when its host star was still in he pre-main-sequence phase. The total loss of water was likely less than an Earth's ocean, however \citep{ribasbolmont2016}. 

If Proxima b has retained sufficient water, partly in liquid form on its surface, life could possibly have formed and developed. 
However, any lifeforms that Proxima~Cen~b was or is hypothetically harboring may have had to develop a UV tolerance much higher than organisms with even the highest UV tolerance on Earth (e.g. {\it Deinococcus radiodurans}). Assuming an Earth-like atmosphere, the recently observed super-flare from its host star and its derived flare rate spectrum suggests that no ozone could survive in the atmosphere of Proxima b, ending all hypothetical life on Proxima Cen b with a single super flare \citep{howardtilley2018}.

It is unclear what types of climate the planet's orbit and assumed tidally-locked or resonating spin rate allow, which also depends on the atmospheric composition. This has, for example, been studied in \citet{turbetleconte2016}, with as main conclusion that conditions on the planet may allow for the existence of liquid water on its surface, for example on the dayside for a tidally locked case, with an 1 bar N$_2$ and slightly enriched in CO$_2$ w.r.t. Earth, or all along the equator in the case of a 3:2 resonance between spin and orbital motion (with 1 bar N$_2$ and enriched in CO$_2$).

With Proxima Cen b's actual atmospheric state unknown, we study the case of an Earth-twin in emitted and reflected light assuming a circular orbit with a radius of 0.0485~AU  \citep{escudeamado}. For modeling the planet spectrum, we used the $P$-$T$ structure of Earth as shown in Figure 1.3 in \citet{bigg2004}. Motivated by the values in Table 1.1 in \citet{bigg2004}, we chose very simple abundance models for the planet atmosphere: for water, a uniform volume mixing ratio of 0.5~\% within the troposphere ($P>0.3$~bar); for ozone 0.7~ppm within the stratosphere ($P<0.3$~bar), and for CO$_2$ and CH$_4$, vertically homogeneous volume mixing ratios of 400 (including alien fossil fuel emission) and 1.75~ppm, respectively. We assumed the a D/H value of $2\times 10^{-4}$, similar to that of Earth.

In the left panel of Figure \ref{fig:HDO_prox} the resulting low-resolution ($\lambda/\Delta\lambda=1000$) synthetic emission spectrum (black solid line) is plotted over the observed average Earth emission spectrum (pink solid line) as measured by the \emph{Nimbus~4} satellite between 6.25 and 25~$\mu$m \citep{hanelconrath1972}. The agreement is close enough for the study presented here. The synthetic spectrum has been obtained with \emph{petitCODE} \citep{mollierevanboekel2015,mollierevanboekel2017}. The reflected light spectrum is shown in red, assuming a planetary surface albedo of 30 \%.

The only atmospheric process we consider for the calculation of the reflected light spectrum is attenuation of the stellar light due to scattering and absorption. Atmospheric emission and scattering into the ray that is propagating through the atmosphere is neglected. Hence, we assume that the stellar flux is reflected at the planet's surface, and attenuated on its way to and from the surface. In this case, the reflected flux measured by an observer at distance $d$ from a planet, when viewing its dayside, is
\beq
F_{\rm refl}(\lambda) = 2 A(\lambda)F_*(\lambda) \left(\frac{R_{\rm P}}{d}\right)^2\int_0^1\mu^2e^{-2\tau(\lambda)/\mu}d\mu \ ,
\eeq
where $A$ is the surface albedo of the planet (assuming \rch{a Lambertian surface}), $F_*$ is the stellar flux measured at the substellar point of the planet, $R_{\rm P}$ is the planetary radius, and $\tau$ is the optical depth from the planetary surface to space, parallel to the surface normal vector. See Appendix \ref{app:refl_flux} for a derivation of this expression. For an airless planet with a surface albedo of unity, one recovers that $F_{\rm refl} = (2/3) F_* (R_{\rm Pl}/d)^2$, hence a geometric albedo of 2/3, as expected.

For the stellar flux, a Proxima~Cen-like stellar spectrum was taken from  \emph{PHOENIX} models \citep{hauschildtbaron1999b}. We assumed an effective temperature and radius of 3042~K \citep{segransankervella2003} and 0.1542~$R_\odot$ \citep{kervellathevenin2017}. 

For the high resolution study in emitted light we calculate the spectrum in the same way as in the low resolution case above, this time using our high resolution code described in Section \ref{sect:high_res_code}. The corresponding high-resolution emission and reflection spectra for an Earth-like Proxima~Cen~b are shown in the right panel of Figure \ref{fig:HDO_prox} for the relevant wavelength range of 3.6 to 3.8 $\mu$m. As is evident from the right panel of Figure \ref{fig:HDO_prox}, the reflected flux is favorable over the emitted by a factor $\sim$15. Since this is a function of the albedo, the various available choices of the latter are discussed at below. Moreover, more and deeper lines appear to be visible in the reflected spectrum when compared to the emission spectrum. This can be understood from the fact that in the emission case isothermal regions in the $P$-$T$ structure tend to decrease the line contrasts, whereas in the reflection case stellar light is simply attenuated as it moves through the planet's atmosphere. The full observed planetary spectrum is simply the sum of the reflected and emitted flux, due to the linear nature of the radiative transfer equation.
 
Our SNR prediction, using Equation \ref{equ:SNR_predict}, suggests that HDO can be detected (at 5$\sigma$) if a planet spectrum can be obtained with an SNR per pixel of 0.23 and 0.11 for the emitted and reflected components, respectively. Making use of Equation \ref{equ:SNR_pix_estimate}, we calculate that a single night (10 hr) of ELT \emph{METIS} observations should result in an SNR of $\sim$0.1 per pixel, when observed in reflected light and assuming a stellar flux reduction at the planet position of a factor $f$=1000. This suggests that if Proxima Cen b is an Earth twin and is similarly enriched in deuterium, HDO could be detectable in $\sim$1 night of observations. However, it will have to be seen how well ELT coronagraphy will work at only 1.8 $\lambda/D$, the star-planet separation at longest elongation. Assuming a maximum practical integration time of 10 nights, a minimum star-light suppression of $f=100$ needs to be achieved to detect HDO in Proxima~Cen~b. We note that we neglected the phase illumination of the planet, assuming full visibility of the planet dayside. If only half of the visible hemisphere of the planet is illuminated by the star during observations, the required observing time increases by a factor 4. 

We stress possible limitations of our study. First on the non-exhaustive list of effects that should be included in future studies, is the choice of the surface albedo, which was taken to be 30~\%. This is the bolometric value expected for ice or snow when considering an M-dwarf host star \citep{turbetleconte2016}. For comparison, ice and snow on Earth have an albedo of $\sim 60$~\%, desert sand has $\sim 40$~\%, whereas oceans and vegetation have albedos of about 5 and 15~\%, respectively \citep[see, e.g., Table 2.2 in][]{marshallplumb2007}. Next, one may consider the effect of scattering the stellar light in the atmosphere itself, and \citet{turbetleconte2016} found that the local bolometric albedo for Proxima Cen b may increase from 30~\% (ice/snow) to 50~\% in regions where water clouds may form. This adds the effect of clouds to the list of processes not considered here. Increasing the albedo due to clouds may seem like an advantage, because a higher albedo increases the reflected flux. At the same time, if the majority of the reflection flux stems from water clouds, then the sharp lines of the water vapor opacity will weaken, as the vapor-richest regions are hidden below the clouds. 
Of course, using an Earth atmosphere to model the HDO detectability should be kept in mind as an important limitation, because Proxima Cen b's atmospheric temperature and chemical abundance structure could be very different from what was assumed here.

\section{Summary and outlook}
\label{sect:discussion_summary}

In this paper we have studied the potential of detecting the $^{13}$C$^{16}$O, HDO, and CH$_3$D isotopologues in exoplanet atmospheres using ground-based high-dispersion spectroscopy. For the $^{13}$C$^{16}$O case we considered the dayside emission spectra of hot Jupiters, while \rch{mostly} concentrating on self-luminous planets for the HDO and CH$_3$D isotopologues. In addition, a HDO reflection study was carried out for Proxima Cen b.
In particular the HDO and CH$_3$D isotopologues are interesting, since their detection will lead to constraints on atmospheric D/H values, providing insights to planet accretion histories and possible atmospheric evaporation processes. In addition, for massive substellar objects, constraints on the D/H value may shed light on their possible deuterium burning history, and hence mass.  

We expect that $^{13}$C$^{16}$O will be readily detectable with instruments such as \emph{CRIRES+} on the VLT. Particularly observations in the \rch{2.4}~micron range, in the \rch{first CO overtone band}, will lead to high detection SNRs for $^{13}$C$^{16}$O, with only a few nights of observations. 

Excitingly, HDO will be detectable at 3.7 $\mu$m with the ELTs over a broad range of atmospheric temperatures, \rch{for planets up to 900~K in equilibrium temperature, when assuming D/H values as low as the galactic mean value}. Since methane tends to blanket the HDO features, atmospheres in which methane is \rch{strongly} quenched may be significantly more accessible, potentially even with 10m class telescopes, \rch{in this case for planets up to 1200~K in equilibrium temperature}. If sufficient coronagraphic starlight reduction can be reached with \emph{METIS} on the ELT at 2$\lambda/D$, an HDO detection in Proxima b will be possible, if its atmosphere is water-rich and Earth-like. 

The CH$_3$D isotopologue, the detection of which will also constrain the D/H value in a planet atmosphere, \rch{would be clearly favorable if the sky background emission at $\sim$4.7~microns could be neglected. However, due to sky background, only the ELT will likely detect \ce{CH3D}, for planets with equilibrium temperatures below 600 K. Also for irradiated or transiting planets in the super-Earth regime, if in the solar neighbourhood, a \ce{CH3D} detection is likely possible.}

Isotopologues will soon be a part of the exoplanet characterisation tools. Here we studied only three isotopologues, $^{13}$C$^{16}$O, HDO, and CH$_3$D, and only consider the signal strength they imprint onto the observed planet spectra. Future studies should also consider different molecules, look into different techniques such as transmission spectroscopy (which we avoided due to the high uncertainty of their cloud and haze properties), and most importantly try to answer how well one can retrieve the actual isotopologue abundance (ratios) from high-resolution observations. The question is whether such observations must be complemented by lower resolution spectroscopy to better constrain, for example, the atmospheric temperature profile. Another unanswered question is whether current line lists have the required precision for carrying out such studies. If line positions of any important atmospheric absorber are wrong, or lines are missing altogether, then the noise arising from pseudo-random line overlap between the observations and models may make the successful isotopologue detection more difficult when using the cross-correlation technique. Striving for the detection of isotopologues in exoplanet atmospheres will challenge the exoplanet modeling community to refine their modeling and retrieval techniques.

\begin{acknowledgements}
We thank our referee Bruno B\'ezard for his many useful comments, which helped improving this paper. P.M. thanks J. Kasting and R. Pierrehumbert for helpful discussions. P.M. also thanks C. Morley and M. Line, whose comments greatly improved the quality of this work. This work benefited from the 2018 Exoplanet Summer Program in the Other Worlds Laboratory (OWL) at the University of California, Santa Cruz, a program funded by the Heising-Simons Foundation. P.M and I.S. acknowledge support from the European Research Council under the European Union’s Horizon 2020 research and innovation programme under grant agreement No. 694513.
\end{acknowledgements}

\bibliographystyle{aa}
\bibliography{mybib}{}

\begin{appendix}
\section{Cross-correlation signal-to-noise}
\subsection{Lines of equal strength}
\label{app:cross_corr_SNR_lines_equ_strength}
Here we derive the signal-to-noise characteristics when trying to identify a line species in noisy data. We start with the case of spectral lines which all have the same strength in the spectral data.
The cross-correlation is defined as
\beq
(f*g)(\tau) = \int_{-\infty}^{\infty} f^*(t)g(t+\tau)dt,
\eeq
it is thus the integral of the product of the function $f$ and the function $g$, where $g$ has been shifted to the `left' (i.e. towards more negative $t$ values) by the distance $\tau$. For real valued functions it holds that $(f*g)(\tau)=(g*f)(-\tau)$.

We now consider a planetary spectrum observed at high resolution within a spectral range $[-\nu,\nu]$. Here we assume that we have a perfect model for the noise-free spectrum of the planet, $f(\nu)$. However, the observation will not give $f(\nu)$, it will give the planetary spectrum plus noise $n(\nu)$. The total observed signal is then $o(\nu)=f(\nu)+n(\nu)$. Throughout this section, for clarity, we will describe the noise assuming a Gaussian with a mean value of zero, and we assume that the planetary spectrum is zero outside the lines, while the lines are assumed to be positive. Because we expect the noise to be fully dominated by the planet's host star, assuming Gaussian-distributed noise is appropriate, as the Poisson distribution transitions to a Gauss distribution at large mean values. Assuming that the noise has a constant standard deviation $\sigma$ for every pixel, a single line within the planetary spectrum, of strength $I$, will only be visible if $I\gg \sigma$. Moreover, the signal-to-noise ratio (SNR) of a single line observation will be ${\rm SNR}_{\rm single} = I/\sigma$.

Now, instead, we carry out a cross-correlation between the observation and the model spectrum, across the range where we have data:
\begin{align}
(o*f)(\nu_0) & = \int_{-\nu}^{\nu} \left[f(\nu)+n(\nu)\right]f(\nu+\nu_0)d\nu \\
& = \int_{-\nu}^{\nu} f(\nu)f(\nu+\nu_0)d\nu +\int_{-\nu}^{\nu}n(\nu)f(\nu+\nu_0)d\nu .
\label{equ:sig_noise_crcr}
\end{align}
Here we neglect the complex conjugate notation, as all functions are real-valued. If one has a perfect (in reality: good) model, then the first integral in Equation (\ref{equ:sig_noise_crcr}) will be maximal for $\nu_0 = 0$. We call this value the signal $S$. This is the signal to be extracted by means of cross-correlation. If the planet has a density of $\rho$ lines per frequency interval in the observed spectral region, with all lines of roughly equal strength $I$, then the height of the peak $S$ at $\nu_0 = 0$ can be approximated as
\beq
S = \int_{-\nu}^{\nu} f(\nu)f(\nu)d\nu  \approx  2\nu\rho I^2 \Delta \nu,
\eeq
where $\Delta \nu$ is the approximate line width of the lines within the studied frequency range and $2\nu\rho$ is the number of lines probed across $[-\nu,\nu]$.
The second integral in Equation (\ref{equ:sig_noise_crcr}), which we will call noise $N$, can be approximated by noticing that the integral will be dominated by the locations where $f(\nu+\nu_0)$ is maximal (i.e. where the lines are). The mean value, for arbitrary $\nu_0$ (assuming that the properties of the planet lines stay constant across the range probed by shifting $\nu_0$) can hence be written as
\beq
N = \int_{-\nu}^{\nu}n(\nu)f(\nu+\nu_0)d\nu \approx \sum_{i=1}^{2\nu\rho}I\sigma_i \Delta \nu.
\eeq
Here, $\sigma_i$ denotes an actual sampled value of the noise with standard deviation $\sigma$.
Now, because we assume the noise values to be independent and random (following a Poisson distribution), we find that \rch{the average magnitude of the noise $N$ is}
\beq
\left<N\right> \approx I\sigma \Delta \nu\sqrt{2\nu\rho}.
\eeq
The SNR measured at the peak of the cross-correlation function of model and observation is thus
\beq
\frac{S}{\left<N\right>} = \sqrt{2\nu\rho}\frac{I}{\sigma},
\eeq
and one sees that the SNR grows with the square root of the number of lines being probed.

In reality the data will contain not only the planetary signal, and the photon noise of the observation, but also lines belonging to the planet's host star, telluric lines, or planetary lines not included in the model $f(\nu)$. If these contaminant lines are uncorrelated with respect to the distribution of the lines in the model spectrum, then they merely correspond to an additional noise source. If this is not the case, and there is some non-negligible correlation, then this will be visible in the form of secondary peaks in the cross-correlation function.

\subsection{Lines of varying strength}
\label{app:cross_corr_SNR_lines_var_strength}

If every line $i$ has a different strength $I_i$, then the signal in the first term of Equation \ref{equ:sig_noise_crcr} can be approximated as
\beq
S = \sum_{i=1}^{N_{\rm lines}}I_i^2\Delta\nu ,
\eeq
whereas for the noise it holds that
\beq
N = \sum_{i=1}^{N_{\rm lines}}I_i \sigma_i \Delta\nu .
\eeq
Because the noise values at the $N_{\rm lines}$ line positions are independent, it holds \rch{for the average noise magnitude} that
\beq
\left<N\right> = \sigma \Delta\nu \left(\sum_{i=1}^{N_{\rm lines}}I_i^2\right)^{1/2} .
\eeq
Thus it holds that
\beq
\frac{S}{\left<N\right>} = \frac{1}{\sigma}\left(\sum_{i=1}^{N_{\rm lines}}I_i^2\right)^{1/2}.
\label{equ:snr_ccr_general}
\eeq
Hence, the strongest lines will be most important. Moreover, when evaluating Equation \ref{equ:snr_ccr_general} for estimating the expected signal-to-noise for a given model $f(\nu)=I(\nu)$, and a given noise level $\sigma$, one can simply set
\beq
\frac{S}{\left<N\right>} = \frac{1}{\sigma}\left(\sum_{i=1}^{N_{\rm \nu}}I_i^2\right)^{1/2},
\eeq
that is taking the sum over all frequency points, rather than just the line positions.

\section{Testing the cross-correlation SNR approximation (Equation \ref{equ:SNR_predict}) for trace species}

\subsection{Testing the SNR approximation for HDO}
\label{sect:first_HDO_test}

Here we present full simulations of the HDO detection for the self-luminous, directly imageable planets, as described in Sections \ref{sect:synth_obs_hdo}. 
This is done to validate our approximative SNR calculations (see Equation \ref{equ:SNR_predict}) used for the HDO detection in Sections \ref{sect:hdo_seld_lum} and \ref{sect:proxima_hdo}. `Full simulation' here means that we included telluric absorption, added photon noise, and then carried out a synthetic reduction step, followed by cross-correlating with an HDO model.  

For the full simulation we first obtained synthetic observations as described in Section \ref{sect:synth_obs}. As before, we assumed to record the data as 100 individual spectra, at a resolution of $10^5$, and three wavelength steps per resolution element.\footnote{Assuming 100 individual spectra is not strictly needed, because the planet's orbital motion is no longer used to remove the tellurics, but we kept this parameter constant when transitioning from the model for hot Jupiters to the self-luminous gas giant observations.} Our benchmark calculations are for young, directly imaged planets, and we therefore assumed that the star-light at the planet's position is reduced by a factor of  $f=1000$. Similar to the CO detection study (see Section \ref{sect:detect_CO}), we treated the stellar flux as featureless, assuming any features can be readily removed. For the velocity offset between the telluric absorption and the planet--star system we assumed a value of 30~km/s, which for all systems near the ecliptic can be achieved due to the barycentric motion of the Earth.

Here we will analyse the characteristics of the HDO detection SNR in planets as a function of $T_{\rm equ}$ at a very large signal-to-noise level, and compare this to the prediction made with the approximation enabled by Equation \ref{equ:SNR_predict}, described in Section \ref{sect:synth_obs_red_hdo}. 
We chose to work with a high signal-to-noise, in order to be able to compare the full and approximative methods properly, otherwise most of the cases would simply lead to non-detections, with the SNRs scattering around zero. The calculations here assume that at the location of the star the spectral single-pixel SNR at 2.3~$\mu$m is ${\rm SNR}_*(2.3 \  {\rm \mu{\rm m}})=5000$ per single spectrum (we will evaluate 100 spectra later), leading to ${\rm SNR}_{*}=3317$ at 3.7~$\mu$m (assuming a sun-like host star). Identically to Section \ref{sect:hd179modelobs}, Equation \ref{equ:stellar_SNR_scaling}, we use a stellar flux model to calculate the corresponding stellar SNR$_*$ at 3.7~$\mu$m. Moreover, because we assume that at the location of the planet the stellar flux is lower by a factor 1000, the photon noise is lower by a factor $\sqrt{1000}$. The data reduction process is described in Appendix \ref{sect:est_sig_noise_hdo}.

\begin{figure}[t!]
\centering
\includegraphics[width=0.495\textwidth]{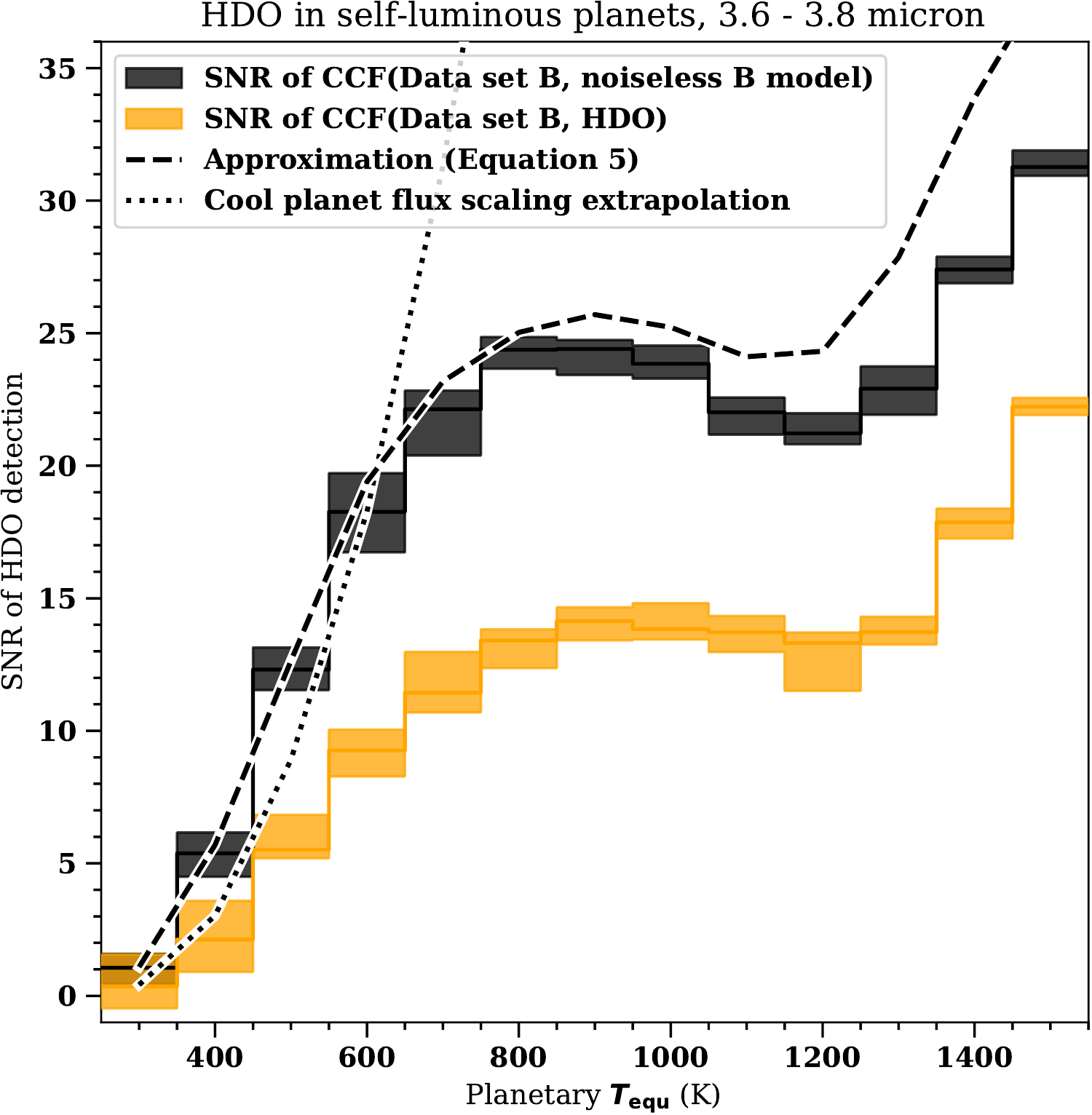}
\caption{SNR of the HDO detection in the spectrum of a directly imaged gas planet as a function of the planetary $T_{\rm equ}$. The HDO/H$_2$O ratio was  $4\times 10^{-5}$. Here we assumed ${\rm SNR}_{*, @ 3.7  \mu{\rm m}}=3317$ per spectrum, a stellar flux 1000 times smaller at the position of the planet, considering 100 spectral observations. Orange boxes: detection SNR when correlating the contaminant removed observation (data set B) with a pure HDO isotopologue template. Black boxes: detection SNR when correlating with  the noiseless, telluric-free model B spectrum. The box height corresponds to the measured 16 and 84 percentiles of the measured SNRs, when running the simulation multiple times. The black dotted line denotes the SNR scaling expected purely from the increasing planet--star contrast as the planet becomes hotter. This assumes that the ratio of the HDO and H$_2$O opacities is not temperature dependent. The black dashed line shows the prediction of the SNR signal following Equation \ref{equ:SNR_predict}, indicating good agreement.}
\label{fig:hdo_tequ}
\end{figure}

The results for the HDO detection SNR are shown in Figure \ref{fig:hdo_tequ}.
We show both the SNRs obtained for cross-correlating the synthetic observation with a pure HDO isotopologue template spectrum (i.e. taking the full self-consistent atmospheric structure but considering only the HDO opacity for the spectrum), and the detection SNR obtained when using the better cross-correlation model $F_{\rm P}-F_{\rm P-HDO}$, which will have the HDO line ratios at correct relative strength. This corresponds to the noiseless and telluric-free model of data set B, described in Section \ref{sect:data_red}. The synthetic observation we correlated the models with was the contaminant corrected data set, analogous to data set B in Figure \ref{fig:data_reduction}. The cross-correlation functions of the 100 synthetic observations we added before calculating the detection SNR. As is evident, using the noiseless and telluric-free model for data set B, instead of the pure HDO isotopologue template spectrum, gives a SNR which is on average better by $\Delta {\rm SNR}\sim 7$.

It can be seen that the detection SNR first increases with temperature, in a way which we find to be linear in the increasing planet--star contrast (black dotted line in Figure \ref{fig:hdo_tequ}) at 3.7~$\mu$m, as expected, as the planet becomes hotter. From $T_{\rm equ}\approx 700$~K on this behavior stops and the SNR stagnates, due to the fact the the HDO opacity is weaker with respect to the H$_2$O opacity at high temperatures. For higher $T_{\rm equ}\gtrsim 1200$~K the HDO detection SNR starts rising again. This is caused by methane, the main absorber in the 3.7~$\mu$m atmospheric window, being less abundant at higher temperatures (see the discussion of Figure \ref{fig:hdo_detect_eq_chem_no_ch4} in Section \ref{sect:hdo_detect_ch4_quench}).

The black dashed line in Figure \ref{fig:hdo_tequ} shows the predicted detection SNR results when using Equation \ref{equ:SNR_predict}. For the spectral signal-to-noise per pixel, used in Equation \ref{equ:SNR_predict}, we evaluated
\beq
(S/N)_{\rm pix} = \frac{c}{\sqrt{1/f+c}} \cdot  {\rm  SNR}_*(3.7~\mu m) ,
\eeq
where $f=1000$ is the flux stellar flux reduction at the position of the planet, and $c$ is the mean planet to star contrast in the wavelength region of interest. 


This makes use of our estimate from Appendix \ref{sect:est_sig_noise_hdo}, namely that the data  reduction process itself is not expected to increase the error of the final spectrum of the target species considerably. One expects Equation \ref{equ:SNR_predict} to follow the detection SNR for cross-correlating the observation with the noiseless and telluric-free model for data set B, and indeed the agreement is very good until equilibrium temperatures of 1100~K. Then Equation \ref{equ:SNR_predict} starts to slightly overpredict the detection SNR. We do not attribute this to the failing of Equation \ref{equ:SNR_predict} at these higher temperatures. Rather, we suspect that the detailed simulation of the HDO detection, which we compare to here, may give rise to additional noise. The noise-free auto-correlation we calculate for the telluric-free data set B is even higher ($\approx 45$ in this temperature range), that is cannot be the reason for the SNR differences we observe here. The SNR approximation method (Equation \ref{equ:SNR_predict}) has thus been verified successfully.

\subsection{Testing the SNR approximation for CH$_3$D}
\label{sect:first_CH3D_test}

\begin{figure}[t!]
\centering
\includegraphics[width=0.495\textwidth]{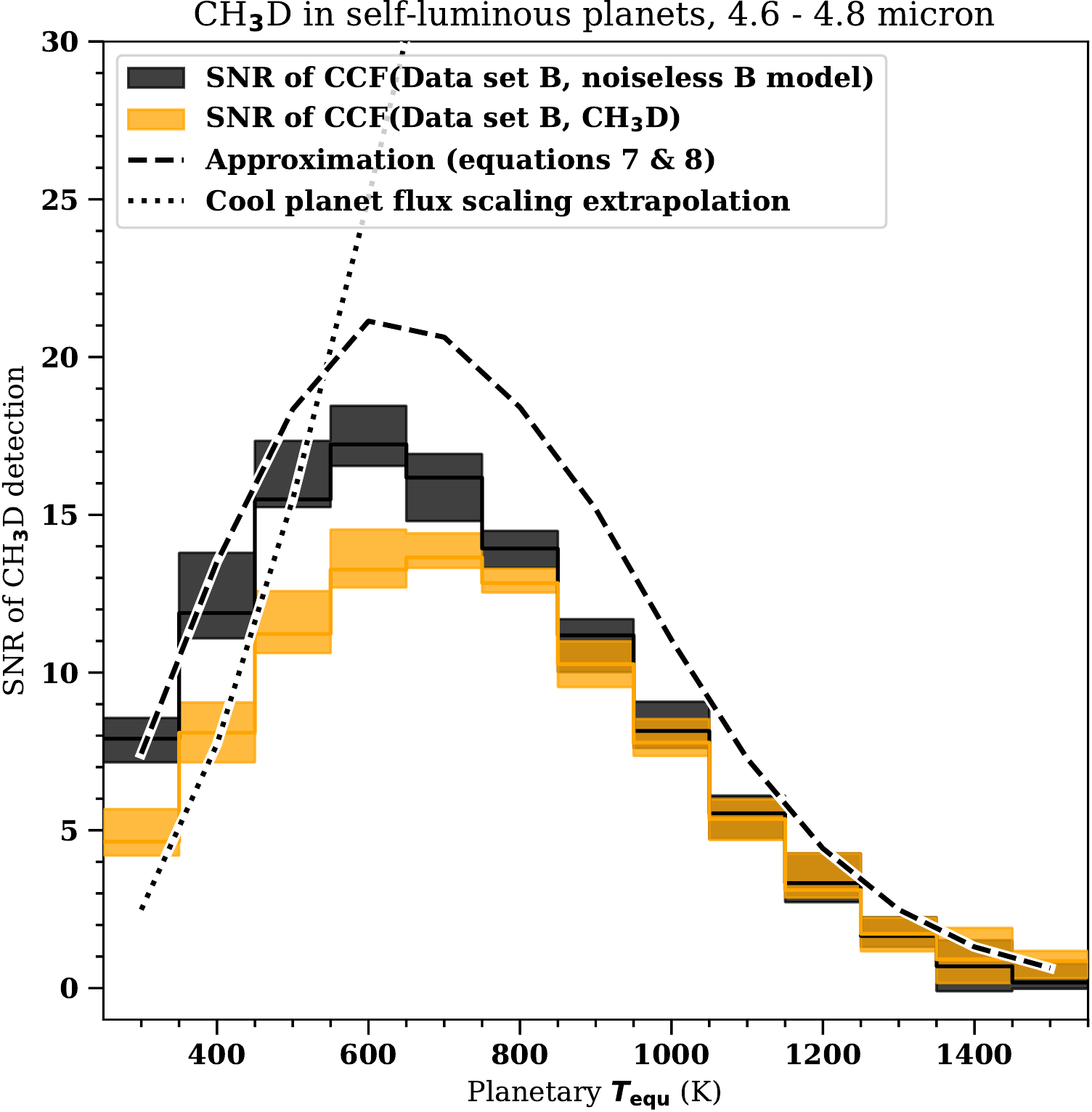}
\caption{Analogous to Figure \ref{fig:hdo_tequ}, but this time testing the SNR-approximation (equations \ref{equ:SNR_predict_ch3d1} and \ref{equ:SNR_predict_ch3d2_bright}) for CH$_3$D.}
\label{fig:ch3d_tequ}
\end{figure}

In Figure \ref{fig:ch3d_tequ} we show the test of the SNR-approximation for CH$_3$D (equations \ref{equ:SNR_predict_ch3d1} and \ref{equ:SNR_predict_ch3d2_bright} in Section \ref{sect:ch3d_self_lum}), analogous to the test shown for HDO in Figure \ref{fig:hdo_tequ}.
One sees that we can reach a good agreement, with differences never larger than 30~\%.

\subsection{Data reduction and estimated error magnitude}
\label{sect:est_sig_noise_hdo}
Here we describe the data reduction used in Appendix \ref{sect:first_HDO_test}, and discuss the resulting magnitude of the planet flux error bars, for a single spectral pixel.

For every of the 100 individual spectra considered here, the observed flux at the Earth's surface is
\beq
F_{\rm surf} =  \mathcal{T}(F_*+F_{\rm P}),
\eeq
where $\mathcal{T}$ is the telluric transmission, $F_*$ the stellar flux at the location of the planet, and $F_{\rm P}$ is the planet's flux (planet and star are velocity-shifted with respect to the telluric transmission). The model input error of a single-pixel and spectrum is supposed to be constant in the approximation here, and given by
\beq
\Delta F_{\rm surf} = \sigma.
\eeq
The telluric transmission model (also see Section \ref{sect:data_red}) will be constructed by observing the star directly, that is not at the location of the planet. We assume that the star, when observed directly, is brighter by a factor $f$. Hence the estimate of the telluric transmission, if assuming that the star's flux without telluric absorption, $F_*$, is perfectly known, is:
\beq
\tilde{\mathcal{T}} = \frac{\left<\mathcal{T}fF_*\right>_{\rm surf}}{fF_*}
\eeq
where $\left<\mathcal{T}fF_*\right>_{\rm surf}$ is the median observed flux in a given pixel, taking all 100 spectra into account. Because the star is brighter by a factor $f$ when observed directly, with the error scaling with $\sqrt{f}$, we find that 
\beq
\Delta\tilde{\mathcal{T}} = \frac{1}{10\sqrt{f}}\frac{\sigma}{F_*},
\eeq
where the factor $1/10$ stems from combining the 100 spectra to get the transmission estimate.
Now, using the measured transmission model, the model flux, taking into account the contaminant species only, is subtracted from the observation (see Section \ref{sect:data_red}), leading to
\begin{align}
F_{\rm rem} & = F_{\rm surf}-\tilde{\mathcal{T}}F_{\rm P-HDO} \\
& \approx \mathcal{T}[F_*+(F_{\rm P}-F_{\rm P-HDO})]
\end{align}
The expected error then works out to be
\beq
\Delta F_{\rm rem} = \sigma\sqrt{1 + \frac{1}{100f}\left(\frac{F_{\rm P-HDO}}{F_*}\right)^2} \approx \sigma .
\eeq
Next, we remove the telluric lines, in order to not be dominated by these in the cross-correlation function:
\beq
F_{\rm remove \ tell} = \frac{F_{\rm rem}}{\tilde{\mathcal{T}}} \approx F_*+F_{\rm P}-F_{\rm P-HDO} .
\eeq
The error works out to be
\begin{align}
\Delta F_{\rm remove \ tell} & = \sqrt{\left(\frac{\sigma}{\tilde{\mathcal{T}}}\right)^2+\left(\frac{1}{10\sqrt{f}}\frac{\sigma}{F_*}\frac{F_{\rm rem}}{\tilde{\mathcal{T}}^2}\right)^2} \\
& = \frac{\sigma}{\tilde{\mathcal{T}}} \sqrt{1+\frac{1}{100f}\left(\frac{F_{\rm rem}}{F_*}\frac{1}{\tilde{\mathcal{T}}}\right)^2} \\
& \approx \frac{\sigma}{\tilde{\mathcal{T}}} \sqrt{1+\frac{1}{100f}} \\
& \approx \frac{\sigma}{\tilde{\mathcal{T}}}.
\end{align}
We note that we implicitly assumed here that the errors introduced by the transmission model estimate $\tilde{\mathcal{T}}$, when obtaining $F_{\rm rem}$, and then $F_{\rm remove \ tell}$, are independent. Only then would it be allowed to carry out two consecutive, and independent error propagation analyses. However, the error introduced by $\tilde{\mathcal{T}}$ is negligible in both steps, and the same would result from a more correct analysis.

Finally, we remove the influence of very strong telluric lines by setting (see Section \ref{sect:data_red})
\beq
F_{\rm final} = \frac{1}{\sigma_{\rm column}}\left(\frac{F_{\rm remove \ tell}}{\left<\left<F_{\rm remove \ tell}\right>\right>}-1\right) ,
\eeq
where $\left<\left<F_{\rm remove \ tell}\right>\right>$ denotes the average over all spectral channels and all 100 spectra, and $\sigma_{\rm column}$ is the standard flux deviation of a given spectral channel, when considering all $F_{\rm remove \ tell}/\left<\left<F_{\rm remove \ tell}\right>\right>$ values for all 100 spectra,  Hence, if there was no HDO line signal, then $F_{\rm final}$ would be simply a flat spectrum around zero, with a standard deviation of 1. One finally gets
\beq
\Delta F_{\rm final} = \frac{1}{\sigma_{\rm column}}\frac{\Delta F_{\rm remove \ tell}} {\left<\left<F_{\rm remove \ tell}\right>\right>}
\eeq
such that the expected signal-to-noise on the HDO-associated planetary flux per spectral pixel is
\beq
\frac{S}{N} = \tilde{\mathcal{T}}\frac{F_{\rm P}-F_{\rm P-HDO}}{\sigma}.
\eeq
The step for obtaining $F_{\rm final}$ (i.e. dividing by $\sigma_{\rm column}$) will effectively remove the wavelength regions of strong telluric lines from the analysis. If one assumes that $\sigma$ itself scales with $\mathcal{T}^{1/2}$, one gets the expected behaviour for photon noise, namely that the SNR of the observation scales with $\mathcal{T}^{1/2}$.


Hence, the reduction and contaminant removal step is not a significant noise source, with the total SNR given simply by the flux difference due to the target species, divided by the observational uncertainties.

\section{Derivation of the reflected flux as a function of surface albedo and atmospheric absorption}
\label{app:refl_flux}

The reflected flux seen by an observer, viewing the dayside of a planet, can be expressed as
\beq
F_{\rm refl} = \int I_{\rm refl}(\Omega)\underbrace{(\mathbf{n}_{\rm P}\cdot \mathbf{n}_{\rm detect})}_{1}d\Omega \ ,
\eeq
where $I_{\rm refl}$ is the intensity of the reflected ray at the top of the exoplanet's atmosphere, $\mathbf{n}_{\rm detect}$ is the normal vector of the detector, and $\mathbf{n}_{\rm P}$ is the direction of travel of the reflected light. Because of the large distances between observer and planet, the angle between the two vectors is negligibly small. Neglecting the atmospheric attenuation, the intensity of the scattered light at the surface of the planet can be found by equating
\beq
F_*(\lambda){\rm cos}(\vartheta) = \pi I_{\rm refl}(\lambda),
\eeq
such that
\beq
I_{\rm refl}(\lambda) = \frac{{\rm cos}(\vartheta)}{\pi}F_*(\lambda),
\eeq
where $\vartheta$ is the angle between the normal vector of the planet's surface and the incoming radiation. \rch{Here a Lambertian surface was assumed}, because for isotropic intensities it holds that $F=\pi I$, which follows from $F = \int I{\rm cos}(\vartheta)d\Omega$, integrated over a solid angle of $2\pi$.

The area of an annulus on the planet's sphere, at an angle $\vartheta$ away from the substellar point, is
\beq
\Delta S = 2\pi R_{\rm P}{\rm sin(\vartheta)} R_{\rm P} \Delta \vartheta \ ,
\eeq
where $R_{\rm P}$ is the planetary radius.
The effective area seen by the observer is equal to ${\rm cos}(\vartheta)\Delta S$. Hence the total reflected flux, using $\Delta \Omega = {\rm cos}(\vartheta)\Delta S/d^2$ for the annulus' solid angle, where $d$ is the distance between planet and observer, is
\begin{align}
F_{\rm refl}(\lambda) & = 2F_*(\lambda)\left(\frac{R_{\rm P}}{d}\right)^2 \int_0^{\pi/2}{\rm cos}^2(\vartheta){\rm sin}(\vartheta) d \vartheta \nonumber \\
& = 2F_*(\lambda)\left(\frac{R_{\rm P}}{d}\right)^2 \int_0^{1}\mu^2 d \mu.
\end{align}
For a variable surface albedo $A(\lambda)$, as well as using that the ray will be attenuated by $e^{-\tau(\lambda)/\mu}$ twice as it travels through the atmosphere, one obtains
\beq
F_{\rm refl}(\lambda) = 2A(\lambda)F_*(\lambda)\left(\frac{R_{\rm P}}{d}\right)^2 \int_0^{1}\mu^2 e^{-2\tau(\lambda)/\mu}d \mu,
\eeq
where $\tau(\lambda)$ is the optical depth of a ray from the planetary surface to space, running parallel to the normal vector of the planetary surface.

\section{Nominal isotopologue abundances}

\begin{table}[t]
\begin{center}
\begin{tabular}{lc}
\hline \hline 
Species  &  Relative abundance \\
\hline
H$_2{^{16}}$O & 99.7 \%\\
HD${^{16}}$O & $4\times10^{-5}$ \\
H$_2{^{18}}$O & $2\times10^{-3}$ \\
H$_2{^{17}}$O & $3.7\times10^{-4}$ \\
HD${^{18}}$O & $6.2\times10^{-8}$ \\
HD${^{17}}$O & $1.1\times10^{-8}$ \\
\hline
${^{12}}$C${^{16}}$O & 98.7 \% \\
${^{13}}$C${^{16}}$O & 1.1 \%\\
${^{12}}$C${^{18}}$O & $2\times10^{-3}$ \\
${^{12}}$C${^{17}}$O & $3.7\times10^{-4}$ \\
${^{13}}$C${^{18}}$O & $2.2\times10^{-5}$ \\
${^{13}}$C${^{17}}$O & $4.1\times10^{-6}$ \\
\hline
CH$_4$ & 98.8\% \\
CH$_3$D & $8\times 10^{-5}$ \\
\hline \hline
\end{tabular}
\end{center}
\caption{\label{tab:nom_iso_abund} Default isotopologue ratios used in this study. The values were taken from the \texttt{molparam.txt} file of the HITRAN/HITEMP databases, which are based on the compilation of telluric isotopic abundances by \citet{debievregallet1984}. Because the Earth is 10-fold enriched in deuterium when compared to the mean galactic value, the relative abundances of D-bearing isotopologues were adjusted such that D/H=$2\times 10^{-5}$.}
\end{table}

In Table \ref{tab:nom_iso_abund} we list the default isotopologue ratios assumed in our calculations, if not otherwise noted in the text.

\section{Parameters assumed for the planets in the HDO and CH$_3$D detection studies}

In Table \ref{tab:planet_params_hdo_ch3d} we list the parameters assumed for the planets in the HDO and CH$_3$D detection studies (sections \ref{sect:hdo_seld_lum} and \ref{sect:ch3d_self_lum}).

\begin{table}[t]
\begin{center}
\begin{tabular}{l|cccccc}
\hline \hline 
Name  &  $T_*$ (K) & $R_*$ ($R_\odot$) & $T_{\rm P}$ (K) & $R_{\rm P}$ ($\rj$) & $a$ (au) & d (pc)\\
\hline
51 Eri b & 7331$^{\rm a}$ & 1.45$^{\rm a}$ & 760$^{\rm b}$ & 1.1$^{\rm b}$ & 13$^{\rm b}$ & 29.4 \\
HR 8799 cde & 7193 & 1.44 & 1200 & 1.05$^{\rm c}$ & 17-43 & 39.4 \\
GJ 504 b & 5978 & 1.36 & 544 & 0.96 & $>$27.8$^{\rm f}$ & 17.56 \\
GJ 1214 b-like$^{\rm d}$ & 3250 & 0.22 & 604 & 0.25 & 0.0149 & 7.28$^{\rm d}$ \\
51 Peg b & 5793 & 1.27 & 1260 & 1.4$^{\rm e}$ & 0.0520 & 14.7 \\
\hline \hline
\end{tabular}
\end{center}
\caption{\label{tab:planet_params_hdo_ch3d} Parameters assumed for the planets in the HDO and CH$_3$D detection studies (sections \ref{sect:hdo_seld_lum} and \ref{sect:ch3d_self_lum}). If not otherwise stated, the data shown here have been culled from \url{http://www.openexoplanetcatalogue.com}. References/notes: a: \citet{rajanrameau2017}, b: \citet{samlandmolliere2017} c: the radii of HR~8799~cde are a mean value for the three planets, the radii of which vary between 1 and 1.1~$\rj$. d: here we assume a planet and star similar in size and temperature to GJ~1214~b, but twice as close to the Solar System as the GJ~1214 system: for every transiting GJ~1214~b-like planet, one may expect up to 10 non-transiting ones, hence at least one at half the distance of the GJ~1214 system. e: this planet is non-transiting, hence its true radius is unknown. We assumed 1.4~$\rj$ here. f: \citet{bonnefoyperraut2018}}
\end{table}

\end{appendix}
\end{document}